\documentclass[12pt,preprint]{aastex}
\begin{document}

\title{Boron Abundances Across the ``Li-Be Dip'' in the Hyades 
Cluster\footnote{Based on observations made with the NASA/ESA Hubble Space 
Telescope, obtained at the Space Telescope Science Institute, which is
operated by the Association of Universities for Research in Astronomy, Inc.,
under NASA contract NAS 5-26555.  These observations are associated with
program \#HST-G0-12294.}}

\author{Ann Merchant Boesgaard\altaffilmark{1}, Michael
G.~Lum\altaffilmark{1}} 

\affil{Institute for Astronomy, University of Hawai`i at M\-anoa, \\ 2680
Woodlawn Drive, Honolulu, HI {\ \ }96822 \\ } 

\email{boes@ifa.hawaii.edu}
\email{mikelum@ifa.hawaii.edu}

\author{Constantine~P.~Deliyannis}
\affil{Department of Astronomy, Indiana University, 727 East 3rd Street, \\
Swain Hall West 319, Bloomington, IN {\ \ }47405-7105 \\ }

\email{cdeliyan@indiana.edu}

\author{Jeremy~R.~King\altaffilmark{1}}
\affil{Department of Physics and Astronomy, Clemson University, 118 Kinard 
Laboratory of Physics, Clemson, SC {\ \ }29634-0978 \\ }

\email{jking2@clemson.edu}

\author{Marc~H.~Pinsonneault, \& Garrett Somers}
\affil{Department of Astronomy, The Ohio State University, 140 West 18th 
Avenue, Columbus, OH {\ \ }43210 \\ }

\email{pinsonneault.1@osu.edu}
\email{somers@astronomy.ohio-state.edu}

\altaffiltext{1}{Visiting Astronomer, W.~M.~Keck Observatory jointly operated
 by the California Institute of Technology and the University of California.}

\begin{abstract}
Dramatic deficiencies of Li in the mid-F dwarf stars of the Hyades cluster
were discovered by Boesgaard \& Tripicco.  Boesgaard \& King discovered
corresponding, but smaller, deficiencies in Be in the same narrow temperature
region in the Hyades.  With the Space Telescope Imaging Spectrograph on the
Hubble Space Telescope we investigate B abundances in the Hyades F stars to
look for a potential B dip using the B I resonance line at 2496.8 \AA.  The
light elements, Li, Be, and B, are destroyed inside stars at increasingly
hotter temperatures: 2.5, 3.5, and 5 $\times$10$^6$ K respectively.
Consequently, these elements survive to increasingly greater depths in a star
and their surface abundances indicate the depth and thoroughness of mixing in
the star.  We have (re)determined Li abundances/upper limits for 79 Hyades
dwarfs, Be for 43 stars, and B in five stars.  We find evidence for a small
drop in the B abundance across the Li-Be dip.  The B abundances for the four
stars in the temperature range 6100 - 6730 K fit the B-Be correlation found
previously by Boesgaard et al.  Models of rotational mixing produce good
agreement with the relative depletions of Be and B in the dip region.  We have
compared our nLTE B abundances for the three high B stars on either side of
the Li-Be dip with those found by Duncan et al. for the two Hyades
giants. This confirms the factor of ~10 decline in the B abundance in the
Hyades giants as predicted by dilution due to the deepening of the surface
convection zone.
\end{abstract}


\section{INTRODUCTION}

The nuclei of the rare light elements, lithium, beryllium, and boron, are
susceptible to destruction by fusion reactions in the interiors of cool
main-sequence stars.  Of this trio, Li is the most fragile, B the most robust.
The destruction of Li by (p,$\alpha$) type reactions occurs at a temperature
of $\sim$2.5 x 10$^6$ K, while Be is destroyed at $\sim$3.5 x 10$^6$ K.
Inasmuch as B is not destroyed until $\sim$5 x 10$^6$ K, it serves as a probe
to even deeper layers inside stars.  The determination of the degree of
depletion of Li, Be, and B in stellar atmospheres serves as a monitor of
internal stellar processes and kinematics.  The surface content of these
elements provides information about the redistribution of matter and the
mixing patterns within a star.  Thus, spectroscopic studies of these three
elements expose the otherwise hidden stellar interior.

The stars of an open star cluster provide a perfect laboratory to make
observations of all three elements for this type of investigation because
their stars have a common origin and a known age and metallicity.  The
abundances of Li, Be, and B that were present in the interstellar medium at
the birth of the cluster are what all of the stars would have had {\it ab
initio}.  Since all the stars start with the same abundance of each light
element, the present-day abundance of each then reveals the deficiency pattern
as a function of stellar mass and thus supplies guidance on the internal
mixing processes.  For main sequence stars in a given cluster the effective
temperature is a good surrogate for stellar mass.

The relatively nearby Hyades cluster is $\sim$7 x 10$^8$ yr old and has a
metallicity of [Fe/H] = +0.13 $\pm$0.02 (Boesgaard 1989, Boesgaard \& Friel
1990).  Recent studies have similar results, e.g.~Paulson et al. (2003) find
+0.13 $\pm$0.01, and Maderak et al.~(2013) find +0.13 $\pm$0.01.  See also
Dutra-Ferreira et al.~(2016).  The errors quoted correspond to the error
in the mean.  The abundances of Li in the main-sequence stars in the Hyades
were first done by Herbig (1965) and Wallerstein et al.~(1965).  The fall-off
of Li with temperature in the Hyades {\bf G dwarfs} has been investigated in
several subsequent papers including Duncan (1981), Cayrel et al.~(1984),
Soderblom et al.~(1990), Thorburn et al.~(1993).  The {\bf F dwarfs} in the
Hyades were studied by Boesgaard \& Tripicco (1986) who discovered large Li
depletions in the mid-F dwarfs; those depletions are factors of 100 relative
to stars only 200 K hotter and cooler.  That data set of 20 stars was
augmented by Boesgaard \& Budge (1988) so that some 30 stars outlined the
Hyades F star Li dip.  This ``chasm'' in the Li abundance occurs in the
temperature range 6400 - 6850 K.  In hindsight we can see that there was a
hint of this Li dip in the Hyades F dwarfs in upper limits found for Li in the
Wallerstein et al.~(1965) paper where they could measure only Li upper limits
on their spectrographic plates.

Boesgaard \& King (2002) completed an investigation of Be in 34 Hyades F and G
dwarfs.  They found the Be counterpart to the Li dip in the {\bf F dwarfs}.
There is no Be counterpart to the Li deficiencies in the {\bf G dwarfs}; this
had been first indicated in a study by Garcia Lopez et al.~(1995) of Be in
four cool Hyades dwarfs.  The Be-dip in the Hyades F stars is not as deep as
the Li dip; Be detections go down to a factor of 7 below the meteoritic value
of 1.42 (Grevesse \& Sauval 1998).

This Li-Be dip contradicts the predictions of standard stellar evolution
theory.  The surface convection zones (SCZ) in mid-F stars are far too
shallow for light element depletion to occur through convective mixing alone.
The standard theory ignores a number of conditions that could affect the light
element abundances, such as rotation, diffusion, mass loss, turbulence,
internal gravity waves, etc.  These effects could result in additional mixing
below the SCZ.  The more observational information we have for all three light
elements, the better we will be able to understand the internal mixing
processes and then build more sophisticated stellar evolution models.

Studies of Li abundances in clusters younger than the Hyades, such as Pleiades
and $\alpha$ Per, found only a small, if any, dip in Li (Pilachowski et
al.~1987, Boesgaard et al.~1988).  The conclusion was that the Li reductions
occurred after the stars were already on the main-sequence; it was not a
pre-main sequence phenomenon.

Explanation for Li depletion was attributed to several causes.  An excellent
discussion of these various mechanisms is given in Stephens et al.~(1997).
Lithium (and other elements) could be carried away in mass loss through
stellar wind mechanisms.  Diffusion between the surface convective and the
radiative zones of the star could reduce the surface abundance.  Pinsonneault
et al.~(1989) investigated a rotating solar model and the effects on
rotationally-induced mixing on Li and Be abundances.  Zahn (1992) found that
such mixing could be applied as an explanation for the sudden depletion of Li
over the narrow range of dip stars.  Using Zahn's models, Charbonnel et
al.~(1994) were able to very closely model the cool, ``red'' side of the dip
for both Li and Be abundances.  Deliyannis \& Pinsonneault (1997) correlated
Li and Be depletions to rotationally induced mixing as an explanation of the
dip.  (Balachandran (1995) showed that post-turnoff stars in M 67 show Li
deficiencies, not the initial Li abundances of the cluster.  Thus the Li had
been destroyed on the main sequence rather than disappearing by nondestructive
diffusion which would have brought up Li atoms from the radiative zone.)

Deliyannis et al.~(1998) and Boesgaard et al.~(2001) have shown that Li and Be
abundances are correlated in F field stars on the cool side of the Li-Be dip
($T_{\rm eff}$ = 5900 - 6650 K).  Boesgaard \& King (2002) discovered that the
Hyades stars in this same temperature range show the same correlation between
A(Li) and A(Be).  The slope -- in both Hyades and field stars -- of this
remarkable logarithmic relationship is +0.36.  Boesgaard et al.~(2004b) found
a similar correlation in the six open clusters and additional field stars.
Their figures 8b and 10 show this correlation for the field and cluster stars
in the range $T_{\rm eff}$ = 6300 - 6650 K which has a slope of +0.43
$\pm$0.04.  Their figures 8a and 9 cover $T_{\rm eff}$ = 5900 - 6650 K with a
slope of 0.38 $\pm$0.03 that is well-matched by the predictions of the
rotation models of Deliyannis \& Pinsonneault (1993, 1997) and Charbonnel et
al.~(1994).

From the observations of the Li-Be dip in F stars in the Hyades and other
clusters, the question naturally arises: ``Is there also a dip in B in the
Hyades F stars?''  This is quite possible as deficiencies of B have been found
in F field stars.  Boesgaard et al.~(2005) discovered a a correlation between
Be and B from HST/STIS Cycle 10 observations of 18 stars of which 13 were very
depleted in Be.  From the B observations in that work they discovered that the
F5 V star, HR 107, is deficient in all three elements.  It is deficient in Li
by a factor of at least 650, in Be by 25 or more, and in B by 10.  Here we
report on the abundance of all three light elements in the Hyades cluster F
dwarf stars.  The well-studied Hyades cluster provides the best venue for this
investigation, but the results are important for our understanding of field
stars as well.

\section{OBSERVATIONS AND DATA REDUCTION}

\subsection{HST Boron Observations}

Observations with the STIS instrument on the {\it Hubble Space Telescope} were
made of vB 14 on 2003 September 9 to determine the B abundance as part of a
program to look at B through the Li-Be dip in the Hyades.  The four other
stars in that program were not observed by the time that STIS died.  After the
servicing mission (SM4) restored STIS to life, we received an additional four
orbits to observe B in two stars in the Li-Be dip region of the Hyades: vB 13
and vB 37.  We were able to supplement these observations with spectra
covering the B region of two Hyades dwarf stars in the HST archive.  The log
of these observations is given in Table 1 and the location of the five stars
in the HR diagram is shown in Figure 1.  The values for luminosity and
effective temperature are taken from de Bruijne et al.~(2001).

The data reduction was provided by the standard HST/STIS and GHRS pipelines.
Figure 2 shows a six \AA{} region of the ultraviolet spectra of two of the
stars, vB 14 and vB 65, near the B I line at $\lambda$2496.  These two stars
differ in temperature by almost 900 K which affects the line strength of the B
I line as well as the blending lines.  One can see significant line-crowding
in this region.

\subsection{Lithium and Beryllium Observations with Keck/HIRES}

Observations of 17 Hyades stars were made at Keck I with HIRES (Vogt et
al.~1994) on the nights of 1999 January 17 and 18 UT.  These spectra covered
the range of 5700 to 8100 \AA\ (with some inter-order gaps) and have a
spectral resolution of $\sim$48,000 and 24$\mu$ pixels.  On each night 15
quartz flat field exposures were obtained along with 15 bias frames.
Comparison spectra of Th-Ar were taken at the beginning and the end of each
night.  Three additional spectra were kindly obtained for us by G.H.~Herbig on
the night of 2005 November 23; these were taken with the updated version of
HIRES with the mosaic CCDs with 15$\mu$ pixels.  Quartz, dark, and Th-Ar
frames were also taken.  
The log of the observations of all 20 stars is given in Table 2.  The
positions of these stars in the HR diagram are shown in Figure 3.

The data reduction was done with IRAF\footnote{IRAF is distributed by the
National Optical Astronomy Observatories, which are operated by The
Association of Universities for Research in Astronomy, Inc., under cooperative
agreement with the National Science Foundation.} and followed standard
procedures with bias subtraction, flat field corrections, order extraction,
wavelength calibration, continuum fitting.  Samples of these spectra are shown
in Figure 4.  The spectrum of the coolest of the three stars shows considerable
depletion of Li while the neighboring Fe I lines (6703, 6705 and 6710 \AA) are
increasing in strength with decreasing temperature.

Most of the observations used here for Be were originally reported by
Boesgaard \& King (2002).  We obtained spectra of 10 additional Hyades stars
for Be with Keck I and the upgraded HIRES with the new high-sensitivity UV
chip with 15$\mu$ pixels.  Those observations are also listed in Table 2.  The
quartz flat-field exposures needed for the Be region were 50 s to achieve the
proper signal that far in the UV.  We took seven quartz frames at 50 s, seven
at 3 s and seven at 1 s as needed for the three CCDs.  In addition we took 11
bias frames and Th-Ar exposures at both the beginning and the end of the
night.  The positions in the HR diagram of the new stars as well as the stars
observed earlier for Be are shown in Figure 5.  Data reduction was done with
the MAKEE pipeline and with IRAF.  Examples of two those new spectra covering
a region of 2.2 \AA{} are shown in Figure 6 where the positions of the two
resonance lines of Be II are marked.

\section{STELLAR PARAMETER DETERMINATION}

de Bruijne et al.~(2001) have done a detailed study of the parameters for the
Hyades stars from the Hipparcos results for 92 high-fidelity stars.  They have
used two calibrations in converting $(B-V)$ and $M_V$ to log $T_{\rm eff}$
and log $(L/L_{\odot})$.

The recalibration of the Hipparcos data by van Leeuwen (2007) give the
distance to the Hyades of 46.45 $\pm$0.50 pc.  The earlier distance from
Perryman et al.~(1998) was 46.34 $\pm$0.27 pc.  We conclude that there is no
justification for changing the parameters of de Bruijne et al.~(2001) based on
the new calibration of van Leeuwen (2007).

We have determined values for the microturbulent velocity, $\xi$, from the
empirical relationship of Edvardsson et al.~(1993) with dependencies on
$T_{\rm eff}$ and log $g$.  Values for stellar masses were determined by
interpolating our derived $T_{\rm eff}$ along the isochrones in Pinsonneault
et al.~2004, which are specific to the Hyades.  We use values of $v$ sin $i$
primarily from Mermilliod et al.~(2009), but also from Paulson et al.~(2003),
Reiners \& Schmidt (2003), and Kraft (1965) for the Hyades.  We adopt
[Fe/H] = 0.13, as discussed in the introduction.

In Table 3 we list the values we use for $T_{\rm eff}$, log $g$, stellar mass,
$\xi$, $v$ sin $i$, and the depth of the surface convection (SCZ) as a
fractional radius.  The star names are listed as van Buren, HD and HIP
numbers.  For the depth of the SCZ of these stars we have used the models
detailed in Table 2 of van Saders \& Pinsonneault (2012).  We have
interpolated among the mass, He abundance, age, and metallicity.  We
calculated the mass of our Hyades sample using temperature values from
Hipparcos data (de Bruijne 2001) and interpolating in Table 2 from
Pinsonneault et al. (2004) which is specific to the Hyades.  We select a He
abundance value of 0.271 $\pm$0.006 from Pinsonneault et al.~(2003), and use
our [Fe/H] value of 0.13.  We select 650 Myr as the Hyades age as a compromise
between the commonly used value of 625 Myr ($\pm$50 My) from Perryman et
al.~(1998), and the slightly older value of 750 Myr ($\pm$100 My) from Brandt
and Huang (2015).  In the case where one of the parameters lies outside of the
van Saders \& Pinsonneault table coverage, we extrapolated using the two
nearest table entries.  If two or more parameters were outside of the range of
the table, we instead interpolated for the SCZ depth directly, using $T_{\rm
eff}$ and depths from similar stars in our Hyades sample.

\section{ABUNDANCE DETERMINATIONS}

Boron abundances were found by spectrum synthesis using the program MOOG(2011
version)\footnote{http://www.as.utexas.edu/$\sim$chris/moog.html} (Sneden
1973).  The line list in the B I region is from Duncan et al.~(1998) with
small modifications as in Boesgaard et al.~(2005).  We excluded lines
from Duncan's original line list that extremely negative log $gf$ values
($<$$-$5) and those with very high excitation potentials ($>$6 ev).  We
relied on the synthesis around the 2496.8\AA~line.  (The wavelengths
given are air wavelengths.)  We also attempted to include synthesis around the
2497.7 \AA~region, but found that blending with the strong iron and
cobalt features in the region prevented any meaningful evaluation of B
abundance from that line.

In the synthesis we adopted [Fe/H] = 0.13 as discussed in the
introduction.  In the models all elements, other than B, are increased by that
same amount.  This is a valid assumption within the 3 \AA{} range surrounding
the fitted region (2495.500-2498.500 \AA{}) for several reasons.  The
abundance of ``iron-peak'' elements (V, Cr, Mn, Co, Ni) plus Al, track with
that of [Fe/H] (Carrera \& Pancino 2011), and the majority of lines in the
full list from Duncan et al.~(1998) (135/185) are from these elements.  Of the
other line sources, Ti and Sc are (possibly) slightly depleted with respect to
iron ($-$0.06 and $-$0.02), respectively, Carrera \& Pancino 2011), however
this minor difference, combined with the relative lack of lines (7/185) does
not contribute to a measurable difference in the synthesis.  The remaining
line sources consist of N, F, P, Cl, and Ge and heavier elements with
insufficiently determined abundances to make accurate adjustments.

Line broadening was calculated by the MOOG synth driver using the ``r''
option, which combines v $sin$ i, macroturbulence, and generic ``Gaussian''
factors. We used the values of v $sin$ i from Table 3 but had no direct
measurement of macroturbulence or of the Gaussian factors (which would include
sources such as instrumental broadening). These two factors were fit as free
parameters during the chi-squared reduction process. The best indicator of
these two factors was the prominent Fe II feature at 2497.3\AA{}, although
other features were also considered.

We determined our synthetic solution using a two-step chi-squared minimization
technique.  Initially, MOOG parameters for line broadening and wavelength and
(normalized) flux offset and scaling, were adjusted manually to minimize the
chi-squared statistic.  The window from 2496.0 to 2497.5 \AA{}, excluding the
area of the 2496.8\AA{} B line was used for this first iteration.  Note that
the B line width varies with the line broadening of a particular star. For
stars with significant broadening, like vB 13, the region excluded in the
first step is approximately 0.6\AA~wide (2496.5-2497.1\AA).  For lines with
lesser broadening (eg. vB 62), the excluded range is slightly less than
0.5\AA.  The second step determined the B abundance which best matches the
observed spectra by minimizing the chi-squared statistic for the region of the
B line; the previously ``excluded'' region.  The final B abundance corresponds
to the abundance which has the lowest chi-squared value for the points within
the B line.

We show the synthesized B spectra in Figure 7 for vB 14, on the hot side of the
Li-Be dip and vB 62 on the cool side of the dip.  The LTE values for A(B) are
2.50 and 2.51, respectively.  In Figure 8 are the syntheses for
two stars in the Li-Be dip: vB 13 and vB 37 at 2.15 and 2.05 respectively

Boron abundances are subject to non-local thermodynamic equilibrium effects,
and we have adjusted our abundance measures, as recommended by Kiselman \&
Carlsson (1996).  Both uncorrected (LTE assumed) and nLTE abundances are given
in Table 4.  When corrected for nLTE effects, the values for A(B)n for vB 14
and vB 62 are 2.73 and 2.55.  The two stars in the Li-Be dip have A(B)n of
2.32 and 2.24 respectively.  

Errors for our B determinations due to the uncertainties in the stellar
parameters are given in Table 5.  Error values were obtained by altering a
single model parameter by the listed amount, and re-running the complete
spectral synthesis in MOOG to determine the resulting change in abundance.
The total uncertainty listed is the square root of the sum of the squares of
the three parameters.  Two of us of us did the B synthesis independently
for 3 of the stars; there was a mean difference of 0.03 dex.

We have determined Li abundances for all 20 stars by spectrum synthesis using
the synth driver in MOOG.  The synthesis for Li is less complex than for
B and there are far fewer lines.  We may have small adjustments in wavelength
and in Gaussian smoothing to make.  This can be seen in Figure 9 where
the fit over the 8 \AA~region is excellent.  The Li abundances are also given
in Table 4.  In addition, we have modified published values (from Boesgaard \&
Budge 1988, Boesgaard \& Tripicco 1986, Cayrel et al.~1984 and Thorburn et
al.~1993) for A(Li) in the other stars to reflect the parameters in Table 3
from de Bruijne et al.~(2001).  Fortunately, A(Li) is dependent only on
the model temperature.  We have found that there is no difference in A(Li)
with log = 4.0 vs.~4.5.  There is no difference with [Fe/H] = +0.1, 0.0, and
even $-$1.0.  The difference in temperature of $\pm$100 K is $\pm$0.08 dex in
A(Li).  There are four stars for which we have new observations with A(Li)
determined here with Li synthesis and for which we redetermined A(Li) with the
equivalent widths from Boesgaard \& Tripicco (1986) and Boesgaard \& Budge
(1988) using the blends driver in MOOG.  The stars are vB 14, 37, 62, 65 and
the $\Delta$A(Li) values are 0.01, 0.00, 0.06 and 0.00, respectively, for
``blends'' minus ``synth.''  Our revised values for A(Li) are given in Table
4.

Takeda et al.~(2013) have determined abundances for Li, C, and O in the
Hyades F and G dwarfs using spectral synthesis.  They also used the de Bruijne
et al.~(2001) stellar parameters derived from {\it Hipparcos} data.  We have
12 stars in common with them for which we both found A(Li) from spectral
synthesis, excluding an additional two for which we both found upper limits on
Li.  The mean difference in our A(Li) minus theirs is $-$0.06 $\pm$0.09.  If
we exclude the two most discrepant stars, vB 26 and 42, among the very coolest
in each of our samples and thus the weakest Li lines, the mean difference is
$-$0.03 $\pm$0.07.  This is excellent agreement.

Beryllium abundances for 34 Hyades F and G dwarfs were done by Boesgaard \&
King (2002) using spectrum synthesis; they were later reanalyzed in Boesgaard
et al.~(2004a) using the later version of MOOG which included Kurucz's UV
opacity edges, the metal bound-free opacity contributions adapted from the
ATLAS model atmosphere code (e.g.~Kurucz et al.~(2011 and references therein.)
We have now readjusted those abundances to reflect the parameters that we have
used here; Be abundances are sensitive to both $T_{\rm eff}$ and $log$
g.  We have synthesized the newly obtained Be spectra, but were unable to get
a reasonable fit for the coolest star, vB 178 at 5235 K.  The Be abundances
for 43 stars are also given in Table 4.

\section{RESULTS}

\subsection{B in the Li-Be Dip}

Figure 10 shows the Hyades Li abundances as A(Li) as a function of $T_{\rm
eff}$ for both Li detections and upper limits.\footnote{Stars that appear in
Table 4 but are excluded from this figure are vB 77 and vB 78 with membership
probabilities of 1\% and 21\% according to Hanson (1975).  We exclude vB 143
and concur with Thorburn et al.~(1993) that it is not a Hyades member.  Also
excluded is vB 124 which is a quadruple system of two SB2 pairs (Griffin et
al.~1985, Turner et al.~1986, Tomkin \& Griffin 2007).}  The drop in Li
in the temperature range 6350 to 6800 K is the Li dip where the values for
A(Li) span nearly two orders of magnitude.  The decline in A(Li) with
decreasing temperatures begins near 6000 K.  Superimposed on this plot are the
nLTE B abundances, A(B)n, on the same vertical scale as the Li abundances.
Both A(Li) and A(B) are plotted with relative to their respective meteoritic
abundances.  There are two stars, vB 13 and vB 37, with B abundances that are
in the Li-dip region.  Both stars appear to be depleted in B.

In Figure 11$^5$ we show the Hyades B and Be abundances on the same vertical
scale as in Figure 10.  They are also plotted relative to their respective
meteoritic abundances.  The Be dip is less pronounced than the Li dip, but
covers almost an order of magnitude.  Although there is a decline in
A(Li) in G dwarf stars cooler than 6000 K, there is no such decline in A(Be).
See Figure 10.  The two stars in the Li dip, vB 13 and vB 37, appear to be
depleted in B with respect to Be as well as with respect to Li.
 
The best mechanism for the decreased values for Li and Be on the cool side of
the Li-Be dip appears to be rotationally-induced mixing at the base of the
SCZ, which mixes Li and Be nuclei down to the region of critical
temperatures for destruction as discussed in the Introduction ($\S$ I).

Figure 12 plots the rotational velocities as v $sin$ i from Table 3 against
their T$_{\rm eff}$ values for our Hyades sample.  B\"ohm-Vitense (2004)
characterized the transition from the most rapidly rotating stars, at T$_{\rm
eff}$ around 7000 K and above, to the slower rotators at 6000 K and below, as
a ``step'' function.  Even with the uncertainty due to $sin$ i, our data
show a steep slope over the temperature region of the Li-Be dip as first noted
by Boesgaard (1987).  The cooler stars have apparently spun down to v $sin$ i
values of $\sim$5 km s$^{-1}$ typically, by 6100 K. 

Over this same range in temperature, the surface convection zone is deepening
as shown in Figure 13 from the data in Table 3.  The SCZ is extremely shallow
in stars hotter than $\sim$7000 K and by $\sim$5500 K it extends over the top
30\% of the stellar radius.  Both the apparent ``spinning down'' of stars
around the dip and the increase of the depth of the convection zone provide
additional information that the depletion of the light elements in this regime
occurs by rotational-mixing mechanisms.  Our measurement of a smaller B
depletion over that of Be and Li in the ``dip'' continues to conform best with
models which incorporate rotationally induced mixing.  We note that in four
Hyades Am stars hotter than 7500 K Burkhart \& Coupry (1989) found no Li
depletion.

Detailed mathematical models of the hydrodynamic processes in the ``dip''
stars by Garaud \& Bodenheimer (2010) used ``gyroscopic pumping'' to model
deep mixing, down the rotation axis, and into regions where all three light
elements can burn. Their Figure 11 shows a diagram of these flows.  While the
process model accounts for Li and Be, it does not specifically deal with B.
However, even without access to a working simulation, their models appear to
predict a lesser, but non-zero depletion of B in the dip stars.  Specifically
we note that, due to the deeper radius of B burning and the linear (log)
scaling of the circulation timescale to that depth, the timescale for B
depletion will be longer than those of Li and Be.  See their Figure 12.
Although their model of gyroscopic pumping reproduces the cool side of the
depletion ``dip'' well, it must rely on (outward) diffusion in order to
replenish (overly) depleted Li and Be (and we assume B) on the hot side of the
dip.  However, they produce a reasonable model for surface replenishment
through a diffusion from the radiative to the convective zone, which brings
the models well in line of the observations of both the Li and Be
non-depletion on the hot side of the dip.

\subsection{The Be-B Correlation}

Of the three elements, Li, Be, and B, it is Li that is the most fragile; the
nuclear reactions that destroy Li occur at temperatures as cool as $\sim$2.5 x
10$^6$ K.  The nuclei of Be atoms can survive at deeper depths where the
temperatures are near 3.5 x 10$^6$ K.  The Li observed in the stellar
photosphere may be severely depleted while the surface Be is unaffected or
mildly depleted.  A correlation between the abundance (and depletions) of Li
and Be in late F- and early G-dwarf field stars was found by Deliyannis
et al.~(1998) and Boesgaard et al.~(2001).

This correlation of Li and Be depletions was well-matched by
rotationally-induced mixing of Deliyannis and Pinsonneault (1993, 1997) and
Charbonnel et al.~(1994).  This was further studied by Boesgaard et
al.~(2004b) in field and cluster stars.  The temperature range where the
depletions are correlated corresponds to the cool side of the Li-Be dip from
the bottom of the Li dip near 6500 K to the Li-peak around 6000 K.  The stars
observed for both Li and Be were 46 field stars and 42 stars from five open
clusters.  In particular, their Figure 12 shows the {\it data} as a line from
the least-squares fit of the observed A(Li) and A(Be) and the model
predictions for three different ages (0.1, 1.7, and 4.0 Gyr) and two different
initial rotational velocities (10 and 30 km s$^{-1}$).  The data and the
theory are well-matched; the theory is slightly offset and is plotted and
A(Be) $-$0.14.  As the stars age they march down that line with less and less
Li and Be with Li decreasing faster than Be.  The inclusion of stars with
different ages revealed the same correlation.  These are slow, but apparently
simultaneous depletions.  The metallicity of the stars is not really relevant
until [Fe/H] $<$$-$0.25.

This work led Boesgaard et al.~(2005) to search for B in F and G dwarf stars
that were depleted in Be.  Such stars would presumably not have detectable Li
as indicated by the Li-Be correlation.  They included five stars with
``normal'' Li and Be abundances.  Inasmuch as B nuclei can survive to deeper
depths where the temperatures are near 5 x 10$^6$ K, they studied B in stars
that were severely Be-depleted.  They found a correlation between the Be and B
(as corrected for nLTE effects) depletions.  As A(Be) goes from 1.4 to $-$0.5,
A(B) declines from 2.4 to 2.0.  Boron is the most robust of the three light
elements.  The slope of A(Be) versus A(B) was 0.22 for the range in $T_{\rm
eff}$ from T = 6100 $-$ 6730 K.  We have overplotted our four cool Hyades
stars on their results for the 14 field stars in Figure 14.  The Hyades stars
match the correlation field stars very well.  The two Hyades stars in the
Li-Be dip with depletions in both Be and B, vB 13 and vB 37, highlight the
reality of the B depletions in the Hyades.  As stars age they deplete both Be
and B as was found for Li and Be.

In Boesgaard et al.~(2004b) we showed how well theory which includes
mixing-enhanced-by-rotation can match the observed plots of A(Li)~vs.~A(Be) in
the temperature range of the Li-Be dip.  With our new data, we present a
matching theory for A(Be)~vs.~A(B) for the observations in Figure 14.  To do
this, we have updated our mixing calculations to include $^{10}$B and
$^{11}$B, with an isotopic ratio of $n(^{10}{\rm B})$/$n(^{11}{\rm B})$ =
0.247 (Lodders 2003) and an initial abundance A(B) equal to the proto-solar
value of 2.6.  Our models track the dominant $^{10}$B($p,\alpha$)$^7$Be and
$^{11}$B($p,\alpha$)$^8$Be reactions, with temperature-dependent rates from
Caughlan \& Fowler (1988) and updated astrophysical S(0)s from the NACRE-II
compilation (Xu et al. 2013). With these additions, we computed mixing models
with masses in the range of 1.15--1.45$M_{\odot}$, and initial angular momenta
log(J) = 50.5 (fast), 50.1 (medium), and 49.7 (slow), where the units of
J are g cm$^{-2}$ s$^{-1}$.  The corresponding values of rotational velocities
are dependent on age, mass etc., but at 700 Myr at 1.25 M$_\odot$ (near the
center of the Li-Be-B dip) they are approximately 15.7, 13.1 and 8.5 km
s$^{-1}$.

Our mixing models are compared to the data in Figure 15. The colored lines
show our predicted A(Be)~vs.~A(B) at 2~Gyr.  These abundance tracks overlap
one another and share the same approximate slope, demonstrating that the
predicted ratio of Be to B depletion is constant regardless of mass, age, and
initial rotation conditions. Our predicted correlation between Be and B
depletion nicely matches the A(Be)~vs.~A(B) trend revealed by the data,
supporting a mixing origin for the abundance ratios of these two elements.  We
note that our models over-predict the B abundance of the Hyades members by
about $\sim 0.1$~dex, perhaps reflecting inaccuracies in the B nuclear cross
sections, missing physics in our models, or a true initial Hyades B abundance
lower by $\sim 0.1$~dex.  (This amount of offset is like that of the A(Li)
vs.~A(Be) described in the second paragraph above.)  However, the
relative B abundance difference of $\sim 0.15-0.2$~dex between the lower
mass stars at A(Be)$\sim$1.1 and the higher mass stars at A(Be)$\sim$0.6 is
accurately predicted.  The black star indicates the highest elemental
depletion predicted by 700~Myr, showing that sufficient B depletion by mixing
is predicted at the age of the Hyades, and that the ratio of Be to B depletion
is constant with age. The good agreement between our predictions and these B
abundances bolsters the success of rotational mixing as an explanation for the
cool side of the Li-Be-B dip.

We stress that the consistency of the Be to B depletion ratio between the
field stars, which presumably have a variety of ages, and the Hyades is very
striking, as is the constancy of the Li to Be depletion ratio, that we have
found repeatedly in our previous works.  This Li-Be-B depletion correlation
dictates very fundamental constraints on how stellar interiors mix.  The
agreement with the present models of B/Be depletion and our earlier similar
models of Li/Be depletion (Deliyannis et al. 1997, 1998) suggests that a star
doesn't care how long it takes to reach a particular Be depletion, or how fast
the star was rotating previously: when a particular Be depletion is achieved,
remarkably, unambiguous corresponding depletions of B and Li are also
achieved.

\subsection{Comparison of B in the Hyades Dwarfs with B in the Hyades Giants}

The light-element abundances have provided many tests of stellar evolution
theory.  Iben (1965, 1967) described the principle that Li, Be, and B would be
diluted as the outer convection zone expanded and deepened during the star's
evolution from the main-sequence to the giant branch.  The small layer of
surface Li would be diluted by Li-free matter {bf below} as the convection
zone deepened, resulting in a major effect on the Li abundance visible on the
surface of giant stars.  The somewhat larger outer layer of Be would suffer a
smaller dilution effect.  These predictions were initially tested for Li by
Herbig \& Wolff (1966) and for Be by Boesgaard \& Chesley (1976).  Similarly,
the even greater reserves of B would be even less diluted.

Duncan et al.~(1998) studied B abundances in two Hyades giants as well as one
dwarf, vB 14. They determined that the two giants were deficient in B relative
to the dwarf by a factor of $\sim$10.  That amount of decrease in B was a good
match for Iben's model predictions.  Other models predict even more B
depletion in red giants.  Sackmann \& Boothroyd (1999) show that the
extra-deep circulation that creates $^7$Li, if long-lived, would destroy all
Be and B.  The Hyades red giants show no excess Li, so that process is not
relevant here.  In addition, their calculations show that if a star spends
more than 10,000 years on the red giant branch both Be and B plummet.  (See
their Figure 7.)  The models of Lagarde et al.~(2012) include extra-mixing
from rotation and thermohaline instability which would circulate Li, Be, B
down to deeper layers.  There more B atoms would be destroyed, beyond the
factor of 10 from dilution alone.  However, for the Hyades giants the
observations show that B is diluted as in Iben's models, but not destroyed.

We have B abundances for one star on the hot side of the Li-Be-B dip and two
on the cool side.  Presumably these three stars contain the initial,
undepleted B in the Hyades.  The mean value is A(B)n = 2.54 $\pm$0.12
consistent with the solar/solar system B abundance.  The value for these three
Hyades dwarfs is similar to that found by Duncan et al.~(1998) and confirms
the conclusions of that paper.

\section{SUMMARY AND CONCLUSIONS}

The dramatic drop in Li and Be abundances in the mid-F dwarfs has prompted our
study of the B abundance in the Li-Be dip.  We obtained spectra of 5 Hyades
stars with HST to find B abundances from the B I line at 2497 \AA.  In
addition we have made observations of Li in 20 stars with Keck I and HIRES for
a total of 79 stars with Li abundances in the Hyades.  We supplemented the 33
stars already observed with Keck for Be by Boesgaard \& King (2002) with 10
new one for a total of 43 Be abundances.  

We used the Hipparcos parameters determined for the Hyades stars by de Bruijne
et al.~(2001) and found abundances by spectrum synthesis for all three
elements.  We have produced new plots of the light-element abundances with
temperature, normalizing all the previous results to the new parameters.

In our overplots of the B-temperature profile (scaled to the Li and to the Be
results) we find evidence for a drop in B abundance across the Li-Be dip.  In
the temperature region of approximately 6400 -- 6800 K, A(Li) drops by a
factor of more than 100, while A(Be) drops by a factor of 10.  The drop in
A(B)n is about a factor of 2.5.  We can now refer to the Hyades Li-Be-B dip.

We also find that A(Be) and A(B)n in the four cool Hyades stars conform to the
B-Be correlation found for field stars in that temperature range by Boesgaard
et al.~(2005).  That logarithmic relationship has a slope of +0.22 as A(Be) is
depleted more than A(B)n.  There is a similar correlation between A(Li) and
A(Be) with a slope of +0.36 where Li is more depleted than Be.  Both
relationships show plausible zero-point shifts at the 0.1 dex level which are
suggested, but not required by the data.

The three of our non-dip stars appear to be undepleted B.  The mean of A(B)n
is 2.54 $\pm$0.12 compared to the solar values of 2.60.  The Hyades giants
will have undergone dilution of the three light elements due to the expansion
of the surface convection zone.  We compared this main-sequence ``initial'' B
abundances with that found in two Hyades giants by Duncan et al.~(1998).  The
B in the giants has been diluted by a factor of $\sim$10, in agreement with
the predictions of standard evolution models by Iben (1965, 1967).

\acknowledgements We are grateful to the Keck Observatory support astronomers
for their knowledgeable assistance during our observing runs.  Special thanks
to the late Dr.~George H.~Herbig for obtaining HIRES spectra of three Hyades
stars for which there were HST B observations.  We would like to thank
Drs.~Jeffrey Rich and C.~J.~Ma for their assistance in data reduction.  We are
also grateful to students Jennifer Beard and Edward Lever for their
assistance.  This work has been supported by HST-GO-09886 and HST-GO-12294,
and NSF grants AST-00-97955 and AST-05-05899.

\clearpage
\begin{deluxetable}{rllllllcr}

\tablenum{1}
\tablewidth{0pc}
\tablecaption{Log of the HST Observations}
\tablehead{
&&&\colhead{Spectral}&&&\colhead{Date}& \colhead{Exp}& \\
\colhead{vB} & \colhead{HD} & \colhead{HIP} & \colhead{Type} & \colhead{V} &
\colhead{B-V} & \colhead{(UT)} & \colhead{(sec)} & \colhead{Instrument} 
}
\startdata
13  & 26345 & 19504 &F6V & 6.576 & 0.429 & 2011 Sep 16 & 9752 & STIS/E230M \\
14  & 26462 & 19544 &F4V & 5.699 & 0.327 & 2003 Sep 20 & 1804 & STIS/E230M \\ 
37  & 27561 & 20357 &F5V & 6.578 & 0.421 & 2011 Sep 16 & 6750 & STIS/E230M \\
62  & 28033 & 20712 &F8V & 7.35  & 0.51 & 2003 Mar 15 & 2358  & STIS/E230H \\ 
65  & 28205 & 20815 &F8V & 7.404 & 0.535 & 1995 Sep 12 & 7072 & GHRS/270M \\

\enddata
\end{deluxetable}

\clearpage
\begin{deluxetable}{rllclllrc}

\tablenum{2}
\tablewidth{0pc}
\tablecaption{Log of the Keck/HIRES Li and Be Observations}
\tablehead{
&&&\colhead{Spectral}&&&\colhead{Date} & \colhead{Exp}& \colhead{S/N}\\
\colhead{          vB} & \colhead{HD} & \colhead{HIP} & \colhead{Type} 
& \colhead{V} &
\colhead{B-V} & \colhead{(UT)} & \colhead{(min)} & \colhead{Li I or Be II} 
}
\startdata
\multicolumn{4}{l}{Li Observations} &&&&&\\
\hline
13 & 26345 & 19504 & F6V & 6.58 & 0.429 & 2012 Jan 07 & \phn{3} & 440 \\
14 & 26462 & 19554 & F4V & 5.71 & 0.327 & 2005 Nov 23 & 0.5 & 499   \\
15 & 26736 & 19793 & G3V & 8.09 & 0.658 & 1999 Jan 17 & 10 & 644  \\
17 & 26756 & 19781 & G5V & 9.16 & 0.696 & 1999 Jan 18 & 18 & 820  \\
18 & 26767 & 19786 & G0V & 8.06 & 0.638 & 1999 Jan 18 & 12 & 721  \\
21 & 284253& 19934 & G5V & 9.11 & 0.816 & 1999 Jan 18 & 15 & 586  \\
26 & 27250 & 20130 & G5V & 8.63 & 0.743 & 1999 Jan 17 & 15 & 612  \\
27 & 27282 & 20146 & G8V & 8.46 & 0.715 & 1999 Jan 17 & 12 & 628  \\
31 & 27406 & 20237 & G0V & 7.47 & 0.566 & 1999 Jan 18 & \phn{6}& 729  \\
37 & 27561 & 20357 & F5V & 6.58 & 0.421 & 2012 Jan 07 & \phn{3} & 440 \\
42 & 27732 & 20480 & G5V & 8.85 & 0.759 & 1999 Jan 17 & 20 & 651  \\
62 & 28033 & 20712 & F8V & 7.35 & 0.051 & 2005 Nov 23 & 1.5 & 432  \\
64 & 28099 & 20741 & G6V & 8.12 & 0.657 & 1999 Jan 18 & \phn{9}& 719  \\
65 & 28205 & 20815 & F8V & 7.42 & 0.535 & 1999 Jan 17 & \phn{5}& 669 \\
73 & 28344 & 20899 & G2V & 7.85 & 0.609 & 1999 Jan 18 & \phn{7}& 718 \\
87 & 28593 & 21099 & G8V & 8.58 & 0.743 & 1999 Jan 17 & 15 & 658 \\
92 & 28805 & \nodata & G8V & 8.66 & 0.741 & 1999 Jan 17 & 15 & 647  \\
93 & 28878 & \nodata & G5V & 9.39 & 0.883 & 1999 Jan 18 & 20 & 633  \\
97 & 28992 & 21317 & G1V & 7.94 & 0.634 & 1999 Jan 18 & \phn{7}& 661 \\
113 & 30311 & 22221 & F5V & 7.26 & 0.549 & 1999 Jan 17 & \phn{5} & 683  \\
121 & 30738 & 22524 & F8V & 7.27 & 0.48 & 2005 Nov 23 & \phn{2} & 510  \\
187 & 32347 & 23498 & K0V & 8.99 & 0.761 & 1999 Jan 18 & 15 & 630  \\
\hline
\multicolumn{4}{l}{Be Observations} &&&&&\\
\hline
1  & 20430 & 15304 & F8V & 7.37 & 0.567 & 2014 Dec 27 & 8.3 & \phn80  \\
2  & 20439 & 15310 & G0V & 7.75 & 0.617 & 2014 Dec 27 & 10  & 108  \\
11 & 26051 & 19261 & F3V & 6.06 & 0.397 & 2014 Dec 27 & 1.7 & 105  \\
18 & 26767 & 19786 & G0V & 8.06 & 0.638 & 2014 Dec 27 & 10.5& 109  \\
40 & 27691 & 20440 & G0V & 7.17 & 0.518 & 2014 Dec 27 & 5   & 111  \\
49 & 27835 & \nodata &G0V &8.20 & 0.605 & 2014 Dec 27 & 10.8& \phn98  \\
52 & 27859 & 20577 & G2V & 7.79 & 0.599 & 2014 Dec 27 & 10  & 112  \\ 
73 & 28344 & 20899 & G2V & 7.85 & 0.609 & 2014 Dec 27 & 10  & 113  \\
118 & 30589 & 22422 & F8V& 7.74 & 0.578 & 2014 Dec 27 & 10  & 130  \\
178 & 28258 & 20850 & G5V& 9.08 & 0.839 & 2014 Dec 27 & 25  & \phn79  \\
\enddata
\end{deluxetable}

\clearpage
\tightenlines
\begin{deluxetable}{rrcccccccc}
\tablenum{3}
\tablecolumns{10}
\tablewidth{0pc}
\tablecaption{Stellar Parameters} 
\tablehead{
\multicolumn{1}{c}{} & \multicolumn{1}{c}{} & \multicolumn{1}{c}{} &
\multicolumn{1}{c}{B-V\tablenotemark{a}}
& \multicolumn{1}{c}{$T_{\rm eff}$} & \multicolumn{1}{c}{log g} 
& \multicolumn{1}{c}{Mass} & \multicolumn{1}{c}{$\xi$} 
& \multicolumn{1}{c}{v $sin$ i} & \multicolumn{1}{c}{SCZ} \\
\multicolumn{1}{c}{vB} & \multicolumn{1}{c}{HD\tablenotemark{b}} & \multicolumn{1}{c}{Hip} 
& \multicolumn{1}{c}{(Hip)} 
& \multicolumn{1}{c}{(K)} & & \multicolumn{1}{c}{(M$_{\odot}$)} 
& \multicolumn{1}{c}{(km s$^{-1}$)} & \multicolumn{1}{c}{(km s$^{-1}$)} 
& \multicolumn{1}{c}{base} 
}


\startdata
1 & 20430 & 15304 & 0.567 & 6079 & 4.446 & 1.19 & 1.38 & \phn5.5 & 0.766 \\
2 & 20439 & 15310 & 0.617 & 5894 & 4.481 & 1.13 & 1.19 & \phn6.4 & 0.737 \\
4 & +23 465 & 16529 & 0.844 & 5223 & 4.579 & 0.94 & 0.53 & \phn3.4 & 0.694 \\
6 & 24357 & 18170 & 0.354 & 7000 & 4.299 & 1.56 & 2.31 & 65.8 & 0.978 \\
8 & 25102 & 18658 & 0.417 & 6705 & 4.329 & 1.45 & 2.03 & 54.0 & 0.910 \\
9 & +19 641 & 18719 & 0.705 & 5598 & 4.534 & 1.03 & 0.88 & \phn4.3 & 0.711 \\
10 & 25825 & 19148 & 0.593 & 5983 & 4.464 & 1.16 & 1.28 & \phn6.5 & 0.749 \\
11 & 26051 & 19261 & 0.397 & 6795 & 4.316 & 1.48 & 2.13 & 25.0 & 0.931 \\
13 & 26345 & 19504 & 0.427 & 6660 & 4.336 & 1.43 & 1.99 & 27.2 & 0.900 \\
14 & 26462 & 19554 & 0.360 & 6971 & 4.301 & 1.54 & 2.29 & \phn8.0 & 0.971 \\
15 & 26736 & 19793 & 0.693 & 5757 & 4.506 & 1.08 & 1.05 & \phn5.6 & 0.722 \\
17 & 26756 & 19781 & 0.640 & 5640 & 4.526 & 1.04 & 0.93 & \phn5.0 & 0.714 \\
18 & 26767 & 19786 & 0.657 & 5812 & 4.497 & 1.10 & 1.10 & \phn4.4 & 0.728 \\
19 & 26784 & 19796 & 0.514 & 6291 & 4.403 & 1.28 & 1.61 & 15.9 & 0.813 \\
20 & 26911 & 19877 & 0.400 & 6783 & 4.318 & 1.48 & 2.11 & \nodata & 0.928 \\
21 & 284253 & 19934 & 0.813 & 5297 & 4.572 & 0.95 & 0.59 & \phn3.9 & 0.697 \\
26 & 27250 & 20130 & 0.745 & 5485 & 4.549 & 1.00 & 0.77 & \phn3.6 & 0.705 \\
27 & 27282 & 20146 & 0.721 & 5553 & 4.541 & 1.02 & 0.84 & \phn5.0 & 0.708 \\
31 & 27406 & 20237 & 0.560 & 6107 & 4.440 & 1.20 & 1.41 & 10.2 & 0.771 \\
36 & 27534 & 20350 & 0.441 & 6598 & 4.345 & 1.40 & 1.93 & 65.9 & 0.886 \\
37 & 27561 & 20357 & 0.412 & 6728 & 4.326 & 1.46 & 2.06 & 20.4 & 0.916 \\
38 & 27628 & 20400 & 0.315 & 7188 & 4.290 & 1.62 & 2.47 & 25.0 & 1.000 \\
39 & 27685 & 20441 & 0.677 & 5692 & 4.518 & 1.06 & 0.98 & \phn4.0 & 0.717 \\
40 & 27691 & 20440 & 0.518 & 6275 & 4.406 & 1.27 & 1.59 & 12.2 & 0.809 \\
42 & 27732 & 20480 & 0.758 & 5449 & 4.554 & 0.99 & 0.74 & \phn3.0 & 0.703 \\
44 & 27731 & 20491 & 0.462 & 6507 & 4.361 & 1.37 & 1.84 & 47.6 & 0.866 \\
46 & 27771 & 20492 & 0.855 & 5196 & 4.582 & 0.93 & 0.50 & \phn4.8 & 0.693 \\
48 & 27808 & 20557 & 0.518 & 6275 & 4.406 & 1.27 & 1.59 & 13.3 & 0.809 \\
49 & 27835 & \nodata & 0.605 & 5934 & 4.474 & 1.14 & 1.23 & \phn2.8 & 0.742 \\
51 & 27848 & 20567 & 0.450 & 6558 & 4.352 & 1.39 & 1.89 & 40.3 & 0.877 \\
52 & 27859 & 20577 & 0.599 & 5961 & 4.468 & 1.15 & 1.26 & \phn6.7 & 0.746 \\
57 & 27991 & 20661 & 0.509 & 6310 & 4.399 & 1.29 & 1.63 & 16.3 & 0.818 \\
59 & 28034 & \nodata& 0.555& 6122 & 4.437 & 1.21 & 1.43 & \phn7.1 & 0.774 \\
61 & 28069 & 20693 & 0.509 & 6306 & 4.400 & 1.28 & 1.62 & 22.3 & 0.817 \\
62 & 28033 & 20712 & 0.556 & 6118 & 4.438 & 1.21 & 1.43 & \phn4.8 & 0.774 \\
63 & 28068 & 20719 & 0.651 & 5776 & 4.503 & 1.09 & 1.07 & \phn7.7 & 0.724 \\
64 & 28099 & 20741 & 0.664 & 5735 & 4.510 & 1.07 & 1.03 & \phn5.8 & 0.721 \\
65 & 28205 & 20815 & 0.537 & 6199 & 4.421 & 1.24 & 1.51 & 11.1 & 0.791 \\
66 & 28237 & 20826 & 0.560 & 6107 & 4.440 & 1.20 & 1.41 & 10.2 & 0.771 \\
69 & 28291 & 20890 & 0.741 & 5497 & 4.548 & 1.00 & 0.79 & \phn4.0 & 0.705 \\
73 & 28344 & 20899 & 0.609 & 5924 & 4.475 & 1.14 & 1.22 & \phn7.6 & 0.741 \\
76 & 283704 & 20949 & 0.766 & 5426 & 4.557 & 0.98 & 0.72 & \phn3.0 & 0.702 \\
77 & 28394 & 20935 & 0.526 & 6242 & 4.412 & 1.26 & 1.56 & 20.1 & 0.802 \\
78 & 28406 & 20948 & 0.451 & 6554 & 4.352 & 1.39 & 1.89 & 28.8 & 0.876 \\
79 & 285773 & 20951 & 0.831 & 5254 & 4.576 & 0.94 & 0.55 & \phn3.9 & 0.695 \\
81 & 28483 & 21008 & 0.470 & 6474 & 4.367 & 1.35 & 1.80 & 24.5 & 0.858 \\
85 & 28568 & 21053 & 0.428 & 6656 & 4.337 & 1.43 & 1.99 & 54.0 & 0.899 \\
86 & 28608 & 21066 & 0.472 & 6465 & 4.369 & 1.35 & 1.79 & 30.4 & 0.856 \\
87 & 28593 & 21099 & 0.734 & 5516 & 4.545 & 1.01 & 0.80 & \phn4.9 & 0.706 \\
90 & 28736 & 21152 & 0.420 & 6693 & 4.331 & 1.44 & 2.02 & 39.8 & 0.908 \\
91 & 28783 & \nodata & 0.880 & 5135 & 4.585 & 0.92 & 0.47 & \phn3.6 & 0.692 \\
92 & 28805 & \nodata & 0.755 & 5453 & 4.553 & 0.99 & 0.74 & \phn4.3 & 0.703 \\
93 & 28878 & \nodata & 0.897 & 5095 & 4.574 & 0.91 & 0.43 & \phn6.5 & 0.690 \\
94 & 28911 & 21267 & 0.429 & 6651 & 4.337 & 1.43 & 1.98 & 40.0 & 0.898 \\
96 & 285931 & 21280 & 0.840 & 5216 & 4.580 & 0.93 & 0.52 & \phn8.1 & 0.694 \\
97 & 28992 & 21317 & 0.631 & 5844 & 4.491 & 1.11 & 1.14 & \phn7.6 & 0.731 \\
99 & 29159 & \nodata & 0.867 & 5165 & 4.585 & 0.92 & 0.47 & \phn3.0 & 0.692 \\
101 & 29225 & 21474 & 0.442 & 6593 & 4.346 & 1.40 & 1.92 & 41.0 & 0.885 \\
102 & 29310 & 21543 & 0.597 & 5968 & 4.467 & 1.15 & 1.27 & \phn7.4 & 0.747\\
105 & 29419 & 21637 & 0.576 & 6045 & 4.452 & 1.18 & 1.35 & \phn4.3 & 0.760\\
106 & 29461 & 21654 & 0.655 & 5762 & 4.505 & 1.08 & 1.05 & \phn2.2 & 0.723\\
109 & 284574 & 21741 & 0.811 & 5303 & 4.572 & 0.95 & 0.60 & \phn4.6 & 0.697\\
113 & 30311 & 22221 & 0.560 & 6107 & 4.440 & 1.20 & 1.41 & \phn7.6 & 0.771\\
114 & 30355 & 22265 & 0.720 & 5555 & 4.540 & 1.02 & 0.84 & \phn4.1 & 0.708\\
115 & 284787 & 22350 & 0.843 & 5225 & 4.579 & 0.94 & 0.53 & \phn2.9 & 0.694\\
116 & 30505 & 22380 & 0.833 & 5249 & 4.577 & 1.00 & 0.80 & \phn4.2 & 0.706\\
118 & 30589 & 22422 & 0.578 & 6037 & 4.454 & 1.18 & 1.34 & \phn5.8 & 0.758\\
119 & 30676 & 22496 & 0.563 & 6094 & 4.443 & 1.20 & 1.40 & 14.3 & 0.769\\
121 & 30738 & 22524 & 0.536 & 6202 & 4.421 & 1.24 & 1.51 & 15.9 & 0.791\\
124 & 30869 & 22607 & 0.501 & 6339 & 4.393 & 1.30 & 1.66 & 21.6 & 0.825\\
127 & 31609 & 23069 & 0.737 & 5508 & 4.546 & 1.00 & 0.80 & \phn3.2 & 0.706\\
128 & 31845 & 23214 & 0.450 & 6558 & 4.352 & 1.39 & 1.89 & 34.2 & 0.877\\
142 & 30246 & 22203 & 0.665 & 5731 & 4.511 & 1.07 & 1.02 & \phn6.6 & 0.720\\
143 & 30809 & 22566 & 0.527 & 6239 & 4.413 & 1.26 & 1.55 & 10.0 & 0.801\\
153 & +29 503 & 13806 & 0.855 & 5196 & 4.582 & 0.93 & 0.50 & \phn4.0 & 0.693\\
162 & 26874 & 19870 & 0.705 & 5601 & 4.533 & 1.03 & 0.89 & \phn1.6 & 0.711\\
178 & 28258 & 20850 & 0.839 & 5235 & 4.578 & 0.94 & 0.54 & \phn3.5 & 0.694\\
180 & 28462 & 20978 & 0.865 & 5172 & 4.584 & 0.93 & 0.48 & \phn3.9 & 0.692\\
187 & 32347 & 23498 & 0.765 & 5429 & 4.556 & 0.99 & 0.72 & \phn4.8 & 0.702\\
\enddata
\tablenotetext{a}{For stars with no HIP number the B-V photometry is from
Joner et al.~(2006) except for vB 91 which is from Upgren \& Weis (1977).}
\tablenotetext{b}{For stars with no HD number the name is the BD number}
\end{deluxetable} 

\clearpage
\tightenlines
\begin{deluxetable}{rrcrrc}
\tablecolumns{6}
\tablenum{4}
\tablewidth{0pc}
\tablecaption{Li, Be and B Abundances} 
\tablehead{ 
\colhead{vB} & \colhead{A(Li)} & \colhead{source\tablenotemark{a}}
 & \colhead{A(Be)\tablenotemark{b}}
 & \colhead{A(B)\tablenotemark{c}} & \colhead{A(B)(nLTE)} 
}
\startdata
1 & 2.96 & 1 & 1.39  &\nodata  &\nodata   \\
2 & 2.70 & 1 & 1.44  &\nodata  &\nodata   \\
4 & $<$0.43 & 1 &\nodata  &\nodata    &\nodata \\
6 & 3.29 & 2 &\nodata  &\nodata  &\nodata   \\
8 & $<$1.87 & 2 &\nodata  &\nodata    &\nodata \\
9 & $<$0.96 & 1 & 1.26 &\nodata  &\nodata   \\
10 & 2.73 & 1 & 1.12 &\nodata  &\nodata   \\
11 & 2.59 & 2 & 0.97 &\nodata  & \nodata  \\
13 & $<$1.70 & 3 & 0.61 & 2.15 & 2.32 \\
14 & 3.36 & 4 & 1.25 & 2.50  & 2.73 \\
15 & 2.40 & 4 & 1.35 &\nodata  &\nodata   \\
17 & 1.98 & 4 & 1.45 &\nodata  &\nodata   \\
18 & 2.53 & 4 & 1.42  &\nodata  &\nodata   \\
19 & 3.00 & 2 & 1.07 &\nodata    &\nodata \\
20 & 3.41 & 2 &\nodata  &\nodata  &\nodata   \\
21 & $<$0.50 & 4 &\nodata  &\nodata    &\nodata \\
26 & 1.30 & 4 &\nodata  &\nodata  &\nodata   \\
27 & 1.65 & 4 & 1.44 &\nodata    &\nodata \\
31 & 2.93 & 4 & 1.28 &\nodata    & \nodata\\
36 & 1.63 & 3 &\nodata & \nodata &\nodata   \\
37 & 2.22 & 3 & 0.60 & 2.05  & 2.24 \\
38 & 2.89 & 3 & 1.16 &\nodata  &\nodata   \\
39 & 2.08 & 1 &\nodata  &\nodata  &\nodata   \\
40 & 3.19 & 1 & 1.44  &\nodata  & \nodata  \\
42 & 1.10 & 4 &\nodata  &\nodata  &\nodata   \\
44 & 2.41 & 3 &\nodata  &\nodata  &\nodata   \\
46 & $<$0.53 & 1 &\nodata  &\nodata    &\nodata \\
48 & 3.06 & 3 & 1.30 &\nodata    &\nodata \\
49 & 2.47 & 1 & 1.42  &\nodata  & \nodata  \\
51 & 1.79 & 3 &\nodata  &\nodata  &\nodata   \\
52 & 2.73 & 1 & 1.42:  &\nodata  &\nodata   \\
57 & 2.84 & 3 &\nodata  &\nodata  &\nodata   \\
59 & 2.88 & 2 & 1.17 &\nodata   &\nodata \\
61 & 3.21 & 2 & 1.23 &\nodata    & \nodata\\
62 & 3.12 & 4 & 1.19 & 2.51  & 2.55 \\
63 & 2.44 & 1 & 1.30 &\nodata  &\nodata   \\
64 & 2.30 & 4 & 1.46 &\nodata  & \nodata  \\
65 & 3.10 & 4 & 1.23 & 2.30 & 2.38 \\
66 & 2.76 & 1 & 1.13 &\nodata  & \nodata  \\
69 & 0.88 & 1 & 1.36 &\nodata  &\nodata   \\
73 & 2.75 & 4 & 1.41  & \nodata &\nodata   \\
76 & 1.14 & 1 &\nodata  & \nodata &\nodata   \\
77 & 2.38 & 2 & 1.01 &\nodata  &\nodata   \\
78 & 2.64 & 2 & 0.87 &\nodata  &\nodata   \\
79 & 0.67 & 5 &\nodata  &\nodata  &\nodata   \\
81 & 2.23 & 3 & 0.93 &\nodata  &\nodata   \\
85 & $<$1.83 & 2 &\nodata  &\nodata  &\nodata   \\
86 & 2.38 & 3 & 0.83 &\nodata  &\nodata   \\
87 & 1.25 & 4 & 1.32 &\nodata  &\nodata   \\
90 & $<$1.25 & 2 &\nodata  &\nodata  &\nodata   \\
91 & $<$0.46 & 1 &\nodata  &\nodata  &\nodata   \\
92 & 1.24 & 4 & 1.32 &\nodata  &\nodata   \\
93 & $<$0.10 & 4 &\nodata  &\nodata  & \nodata  \\
94 & $<$0.25 & 3 &\nodata  &\nodata  &\nodata   \\
96 & 0.75 & 1 &\nodata  &\nodata  & \nodata  \\
97 & 2.65 & 4 & 1.22 &\nodata    &\nodata \\
99 & 0.49 & 1 &\nodata  &\nodata  &\nodata   \\
101 & $<$1.07 & 2 & 0.55 &\nodata    &\nodata \\
102 & 2.74 & 1 &\nodata  &\nodata  &\nodata   \\
105 & 2.82 & 1 &\nodata  &\nodata  &\nodata   \\
106 & 2.45 & 1 & 1.38 & \nodata   & \nodata\\
109 & $<$0.74 & 1 & \nodata & \nodata    &\nodata  \\
113 & 2.84 & 4 & 1.18 &\nodata    &\nodata \\
114 & 1.69 & 1 & 1.43 &\nodata    &\nodata \\
115 & $<$0.43 & 1 &\nodata  &\nodata   & \nodata\\
116 & $<$0.48 & 1 &\nodata  &\nodata    &\nodata \\
118 & 2.74 & 1 & 1.44  &\nodata  &\nodata   \\
119 & 2.74 & 1 &\nodata  &\nodata  &\nodata   \\
121 & 3.15 & 4 & 1.28 & \nodata &\nodata   \\
124 & 1.78 & 3 & 0.71 & \nodata &\nodata   \\
127 & 1.33 & 1 &\nodata  &\nodata  &\nodata   \\
128 & 2.24 & 3 & 0.92 &\nodata    &\nodata \\
142 & 2.24 & 1 &\nodata  &\nodata  &\nodata   \\
143 & $<$1.15 & 1 &\nodata  & \nodata   &\nodata \\
153 & 0.83 & 1 &\nodata  &\nodata  &\nodata   \\
162 & 2.16 & 1 &\nodata  &\nodata  &\nodata   \\
178 & 0.67 & 1 &\nodata  &\nodata  &\nodata   \\
180 & 0.60 & 1 &\nodata  &\nodata  &\nodata   \\
187 & 1.38 & 4 &\nodata  &\nodata  &\nodata   \\
\enddata
\tablenotetext{a}
{
1) Thorburn et al.~(1993) equivalent width measures, our new parameters;
2) Boesgaard \& Budge (1988) equivalent width measures, our new parameters;
3) Boesgaard \& Tripicco (1986) equivalent width measures, our new parameters;
4) Li synthesis in this study;
5) Cayrel et al.~(1984) equivalent width measures, our new parameters.}
\tablenotetext{b}{A(Be) from Boesgaard et al.~(2004b) synthesis, corrected 
for our new parameters and for nine of the stars observed for Be here.}
\tablenotetext{c}{All B from synthesized spectra (HST)}

\end{deluxetable}


\clearpage
\tightenlines
\singlespace
\begin{center}
\begin{deluxetable}{rccccc}
\tablenum{5}
\tablewidth{0pc}
\tablecolumns{6} 
\tablecaption{Boron Abundance Errors due to Uncertainties in the Stellar 
Parameters}
\tablehead{
\multicolumn{1}{c}{vB} &
\multicolumn{1}{c}{A(B)} &
\multicolumn{1}{c}{$\Delta$T$_{\rm eff}$} &
\multicolumn{1}{c}{$\Delta$log g} &
\multicolumn{1}{c}{$\Delta$[Fe/H]} & 
\multicolumn{1}{c}{Total} \\
\multicolumn{1}{c}{} &
\multicolumn{1}{c}{} &
\multicolumn{1}{c}{$\pm$75 K} & 
\multicolumn{1}{c}{$\pm$0.20} & 
\multicolumn{1}{c}{$\pm$0.10} & 
\multicolumn{1}{c}{}
}
\startdata
13  & 2.32  & $\pm$0.04  &  $\pm$0.02 & $\pm$0.01 &  $\pm$0.05 \\
14  & 2.73  & $\pm$0.08  &  $\pm$0.02 & $\pm$0.03 &  $\pm$0.09 \\
37  & 2.24  & $\pm$0.07  &  $\pm$0.02 & $\pm$0.01 &  $\pm$0.07 \\
62  & 2.55  & $\pm$0.04  &  $\pm$0.05 & $\pm$0.07 &  $\pm$0.09 \\
65  & 2.38  & $\pm$0.08  &  $\pm$0.07 & $\pm$0.04 &  $\pm$0.11 \\
\enddata

\end{deluxetable} 
\end{center}

\clearpage

\begin{figure}
\plotone{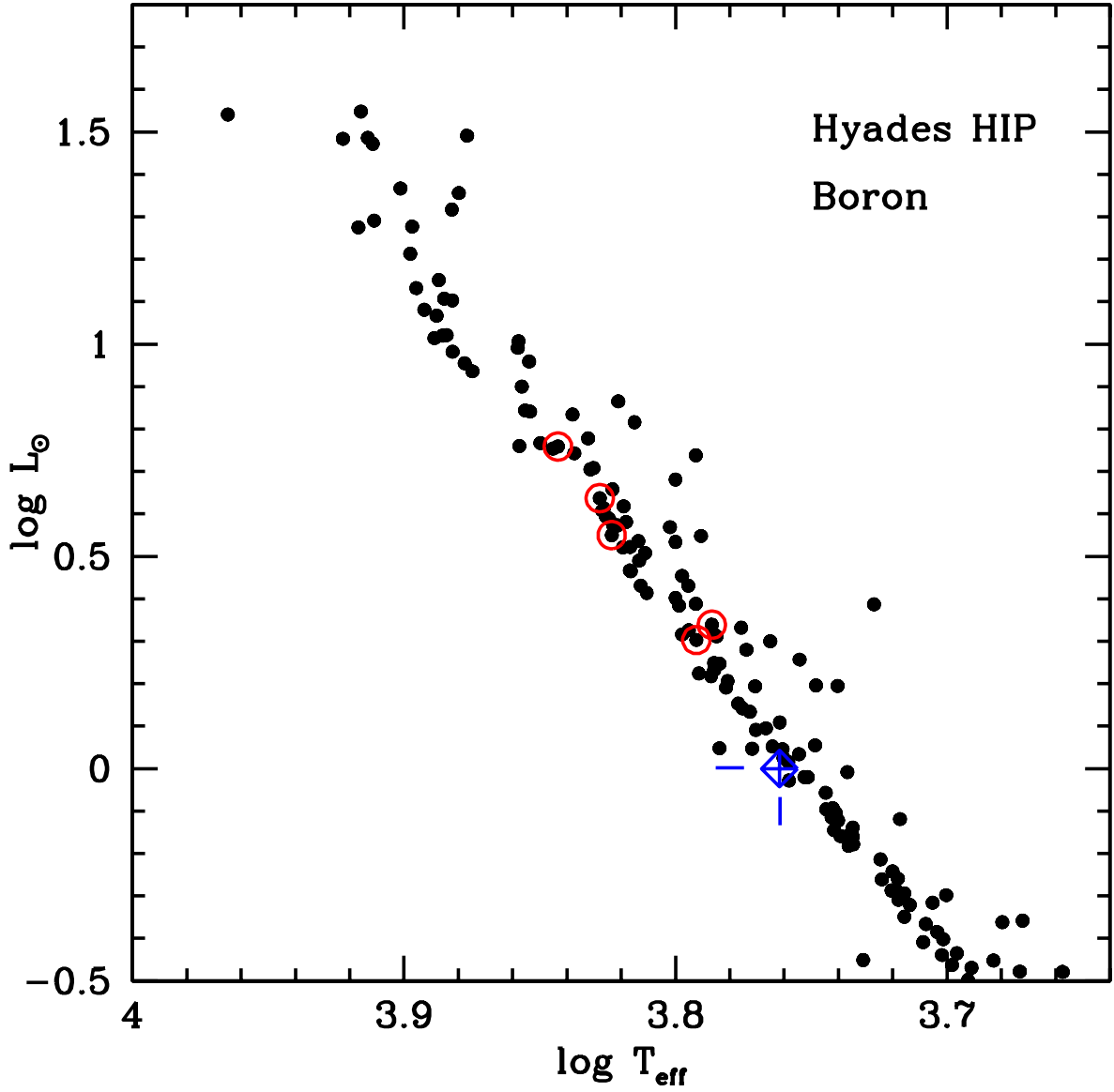}
\caption{The color-magnitude diagram for the Hyades from Hipparcos data
showing the positions of the stars observed for B for this program and from
the HST archives circled in red.  The position of the sun is indicated by the
crossed diamond and further delineated by the short vertical and horizontal
lines in blue.}
\end{figure}

\begin{figure}
\plotone{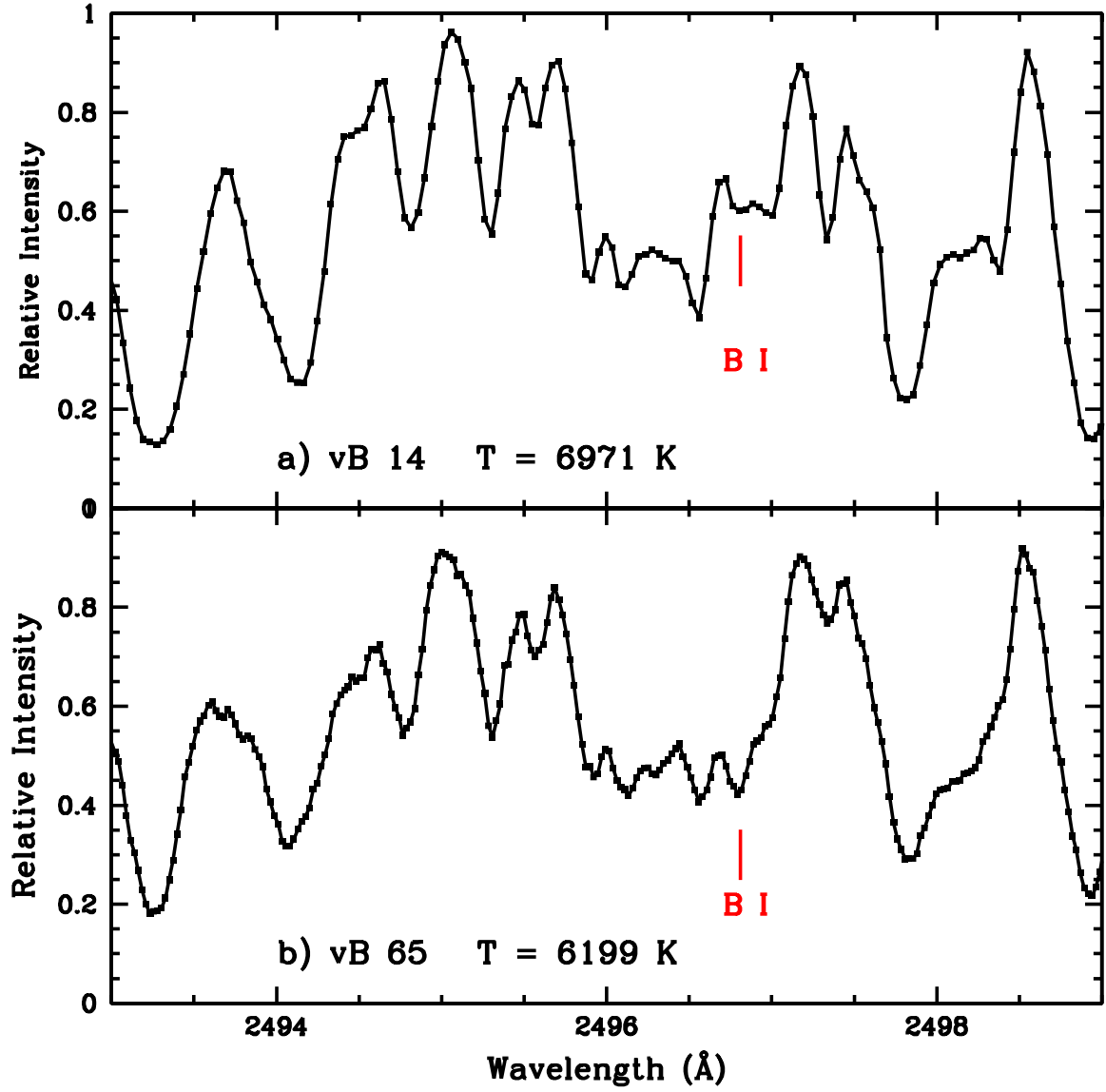}
\caption{The spectra of two of our stars in the B I region, indicating the B I
line that we used in the B abundance analysis.}
\end{figure}

\begin{figure}
\plotone{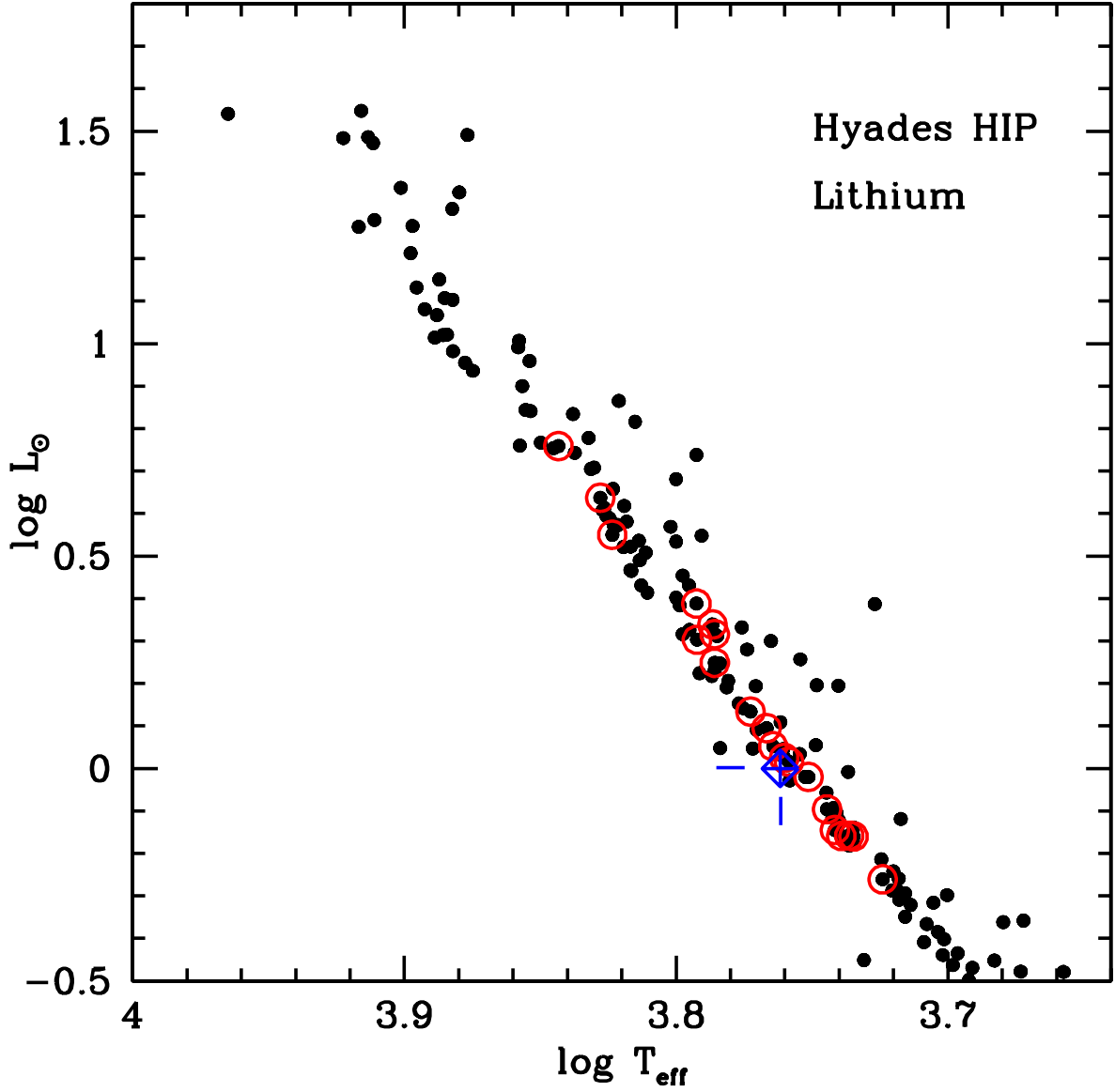}
\caption{The color-magnitude diagram for the Hyades from Hipparcos data.  The
stars newly observed for Li are circled.  As in Figure 1 the position of the
Sun is indicated.}
\end{figure}

\begin{figure}
\plotone{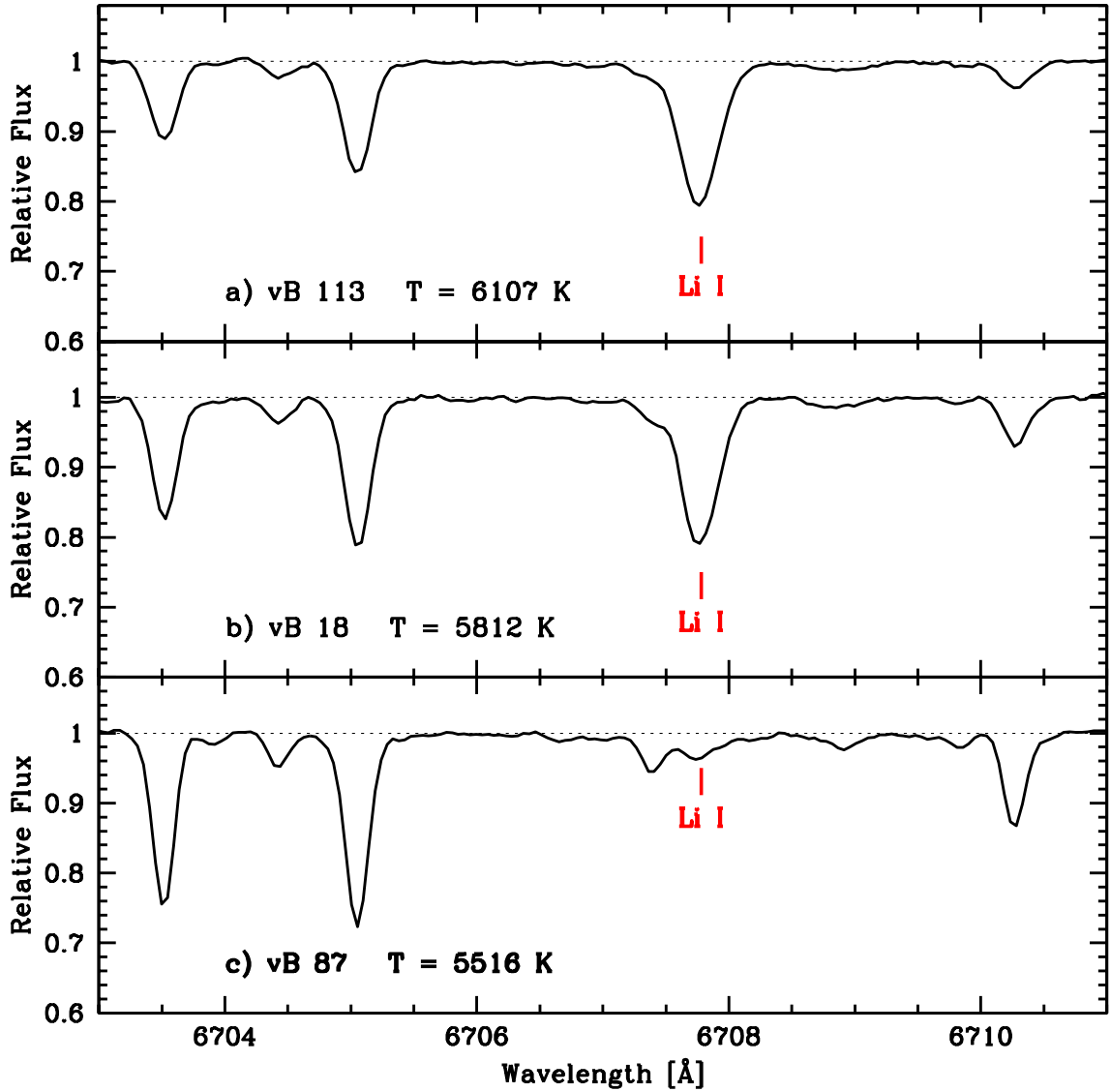}
\caption{A region eight \AA\ in the spectra of three stars near the Li I line
at $\lambda$6707.  The three stars show a range in temperature and the Li I
doublet is obviously much weaker in the coolest star.  It is very depleted in
Li.}
\end{figure}

\begin{figure}
\plotone{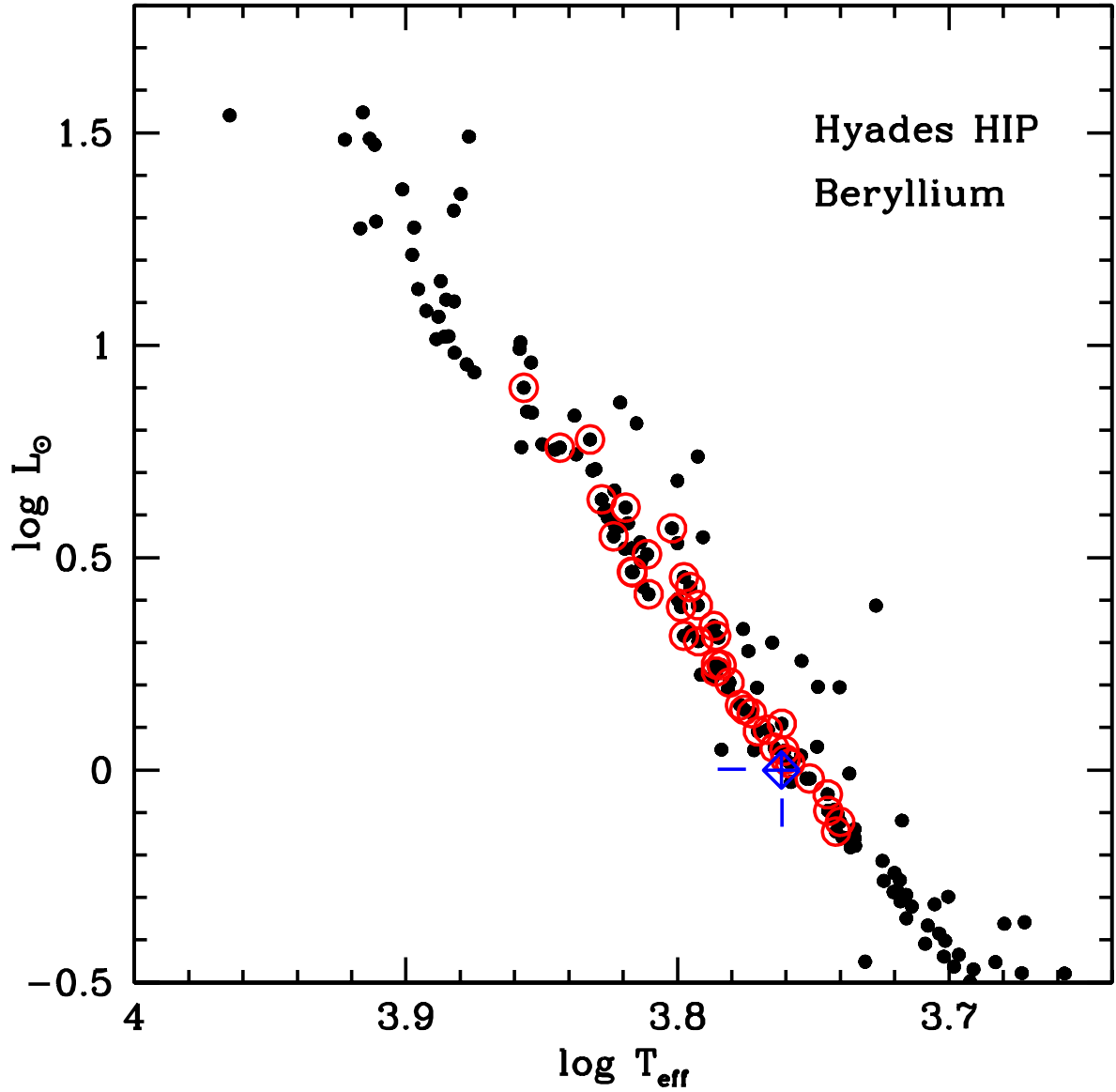}
\caption{The color-magnitude diagram for the Hyades from Hipparcos data
showing the positions of the stars observed for Be by Boesgaard and King
(2002) and in this work.  As in Figure 1 the position of the Sun is
indicated.}
\end{figure}

\begin{figure}
\plotone{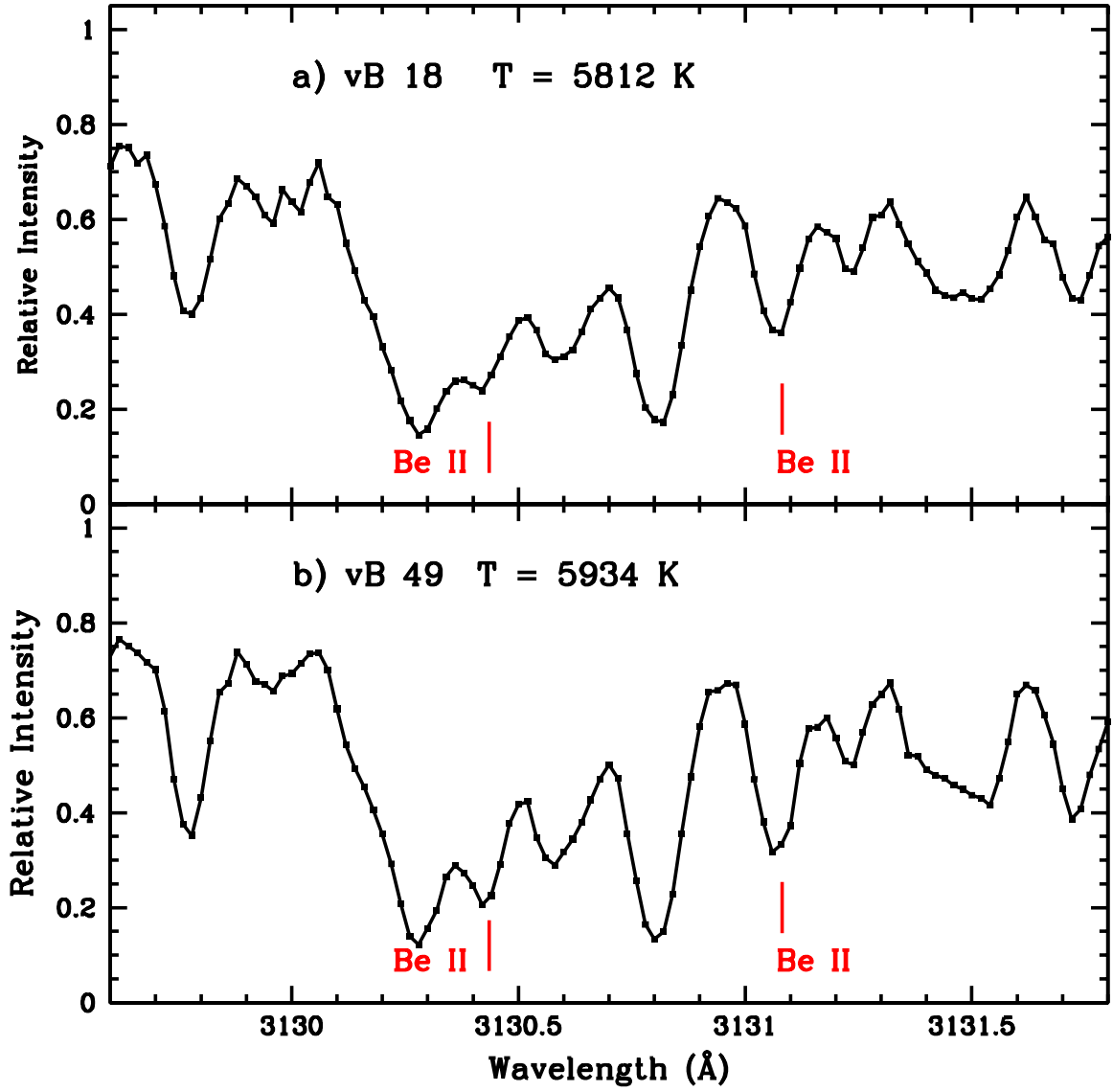}
\caption{The spectra of Be for two of our new observations}
\end{figure}

\begin{figure}
\plotone{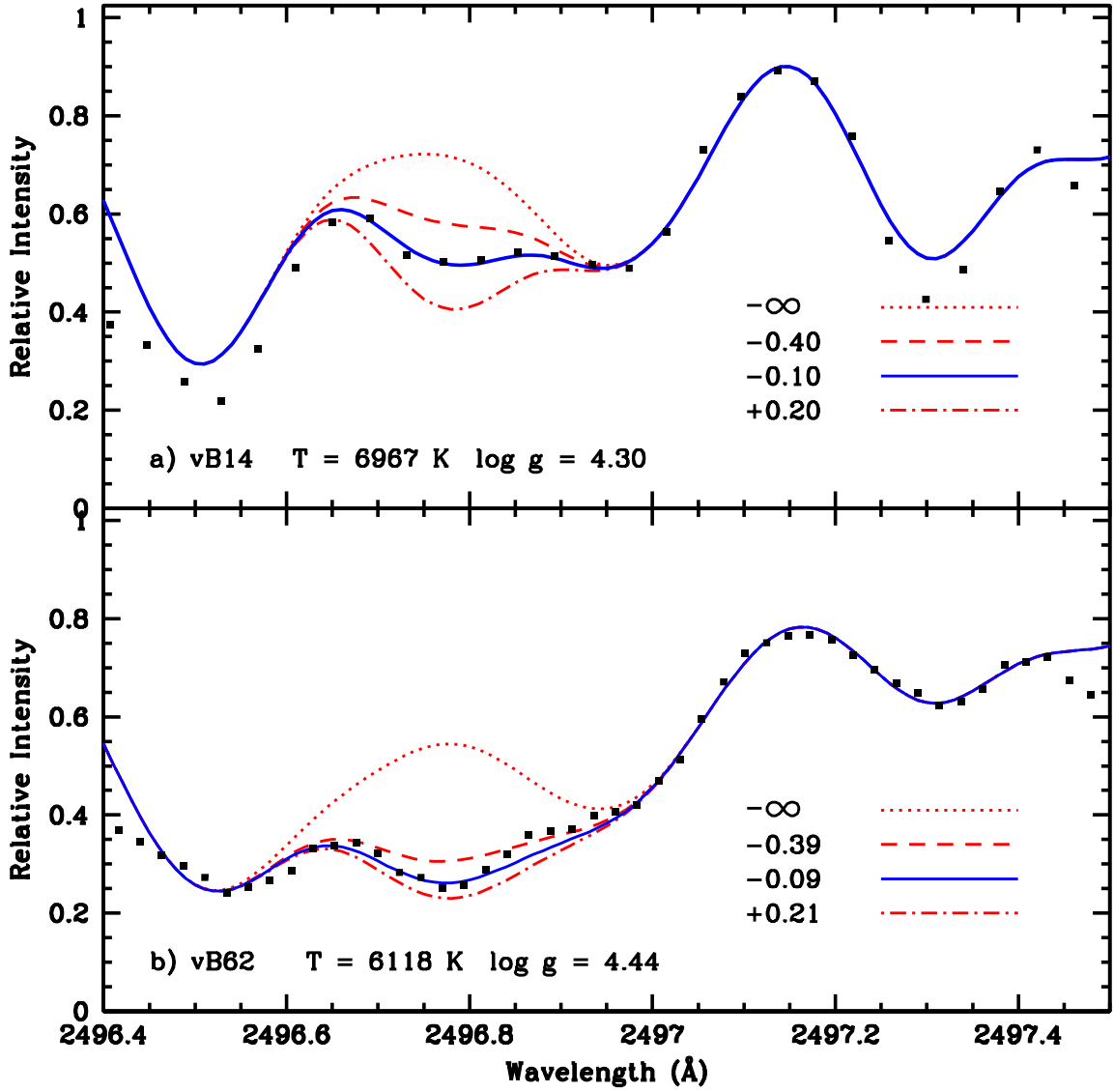}
\caption{The synthesis fits for B in vB 14 and vB 62.  The observations are
the black squares.  The best fit synthesis is shown by the blue line.  The B
abundance of a factor of 2 larger is shown as the red dot-dash line and a
factor of 2 smaller by the red dashed line.  The synthesis with no B is the
red dotted line.}
\end{figure}

\begin{figure}
\plotone{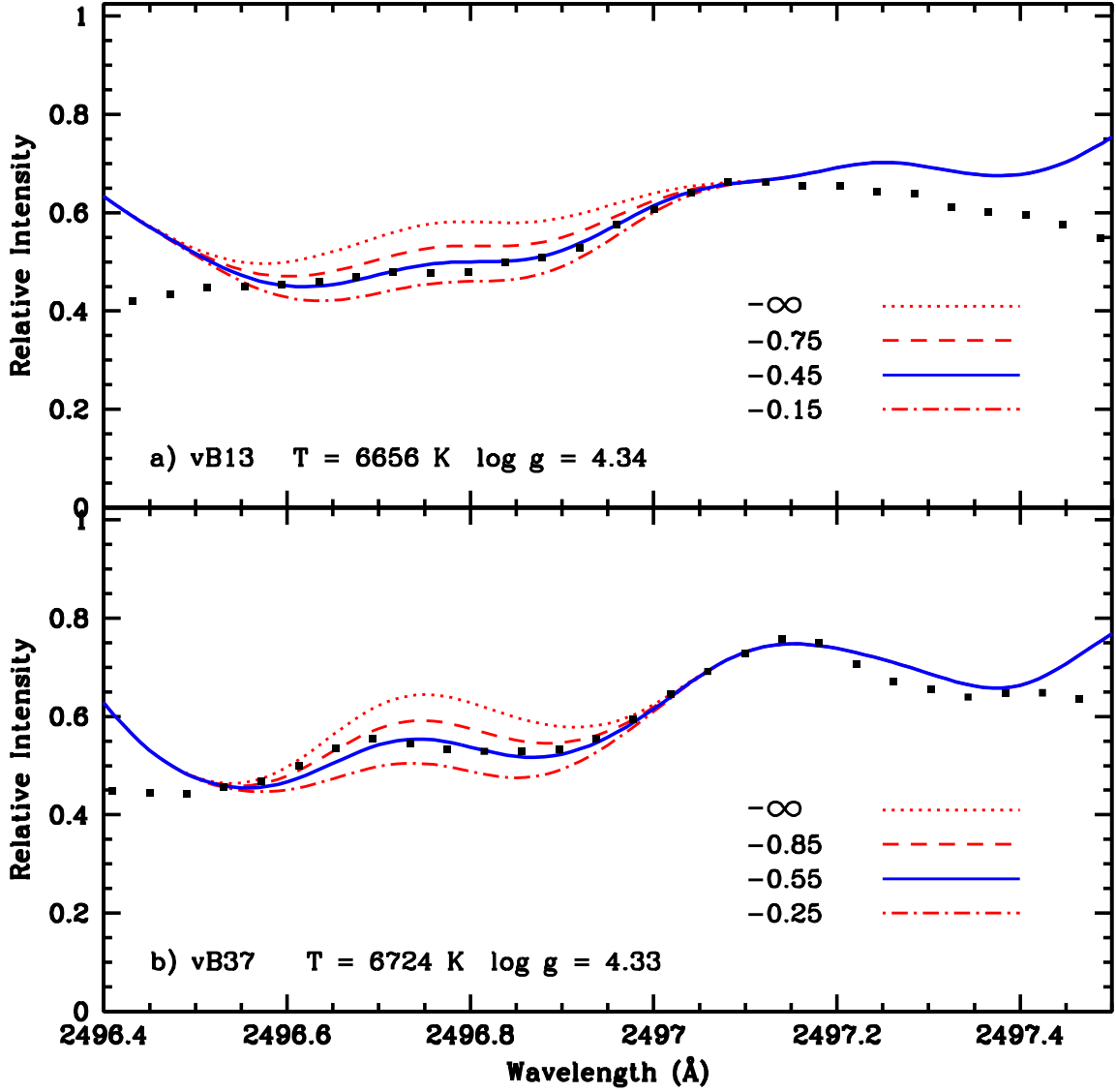}
\caption{The synthesis fits for B in the two Li-Be dip stars, vB 13 and vB 37.
The lines and symbols are the same as in Figure 7.}
\end{figure}

\begin{figure}
\plotone{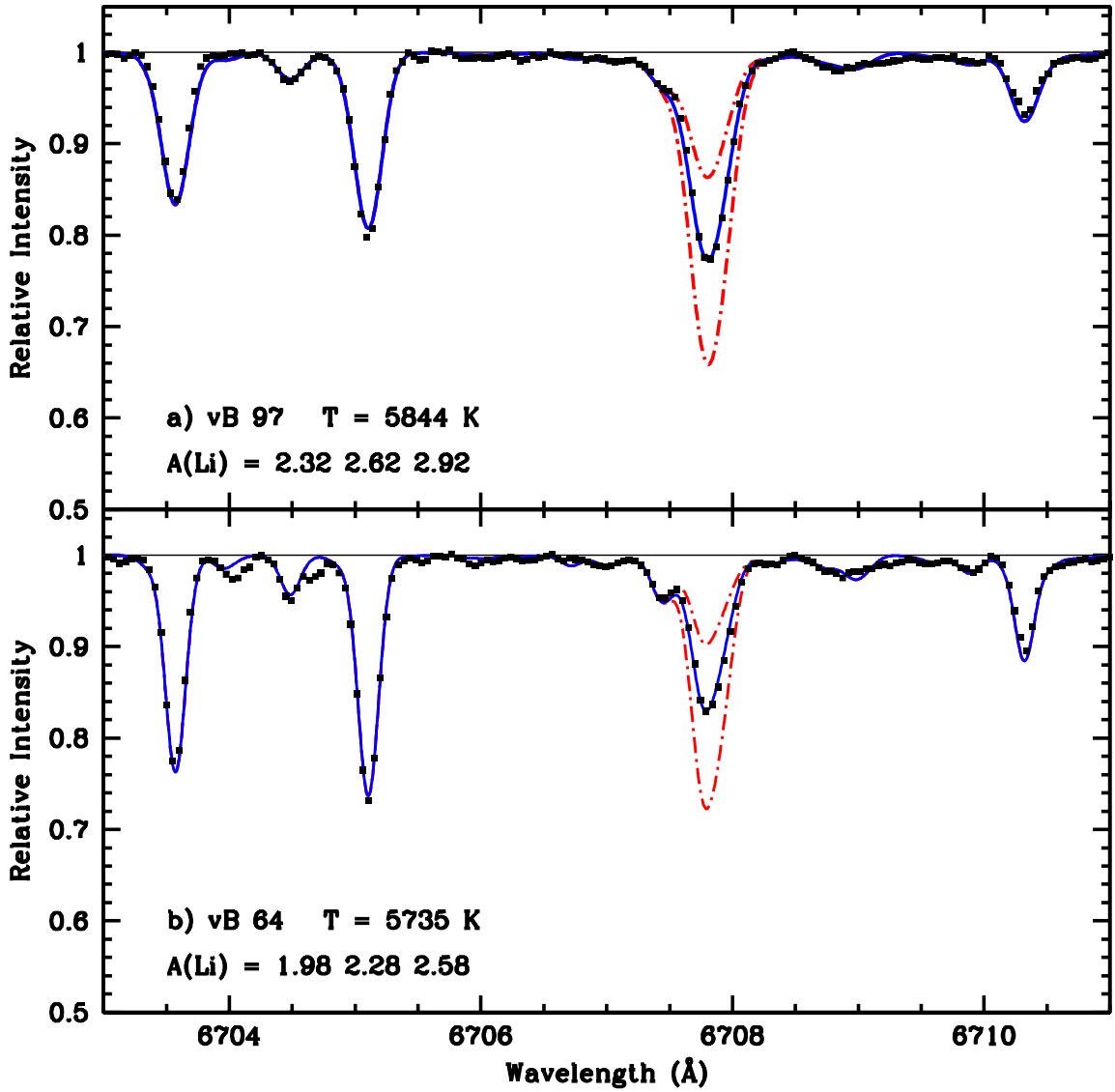}
\caption{Example of the Li synthesis for vB 97 and vB 64. The observations are
the black squares.  The best fit synthesis is shown by the blue line.
Abundances of Li a factor of 2 larger and smaller are shown by the red
dash-dot line.  The position of the continuum is the narrow black line at
1.00.}
\end{figure}

\begin{figure}
\plotone{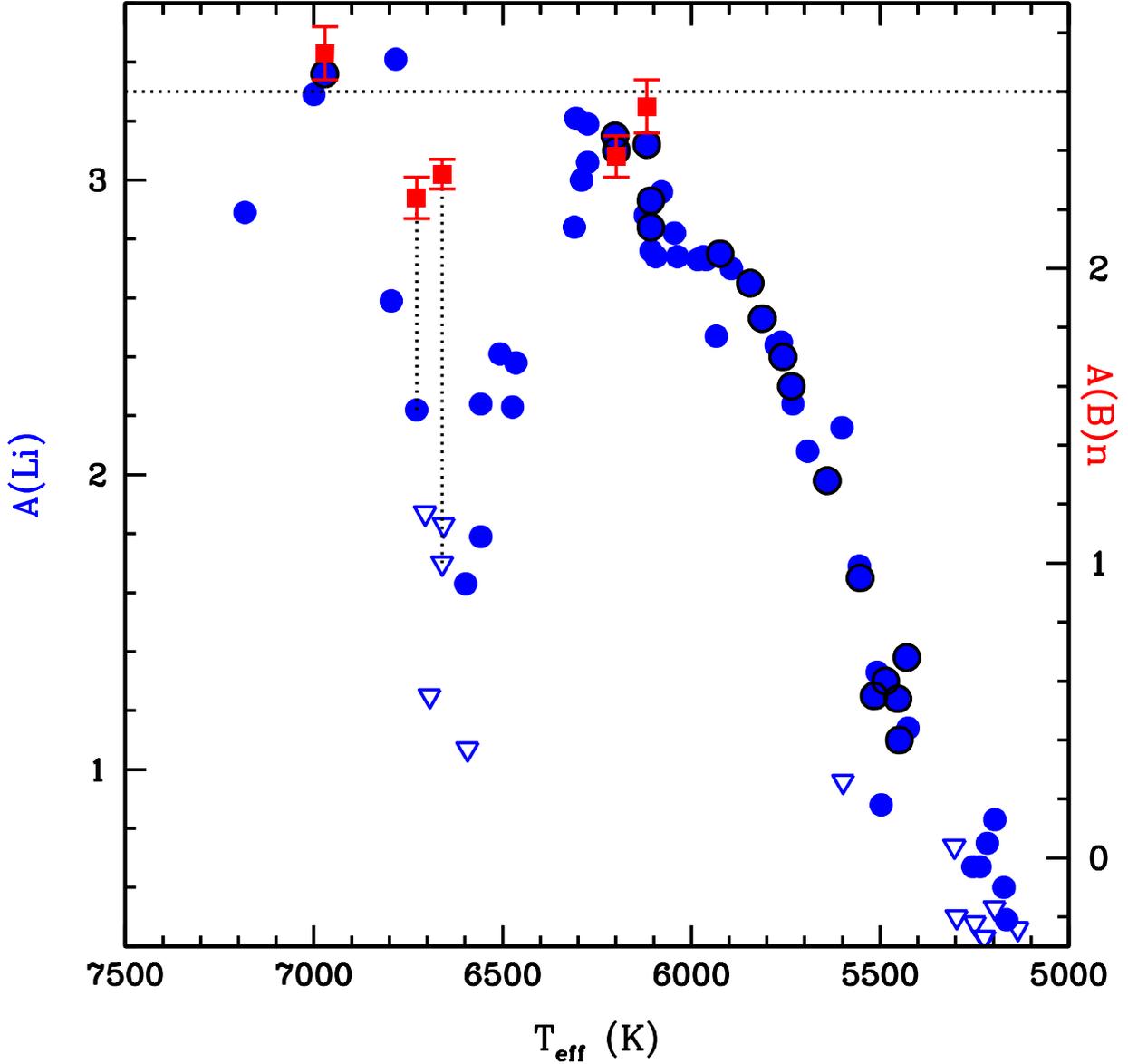}
\caption{The Li-temperature profile for the Hyades F and G stars.  The blue
filled circles are Li detections while the blue open triangles are Li upper
limits.  The Li abundances from the Li observations reported in this work are
circled in black.  Note the pronounced drop in Li abundances in the range 6400
- 6850 K, the Li ``dip.''  The Li abundances also drop sharply in the cooler G
stars with T $<$ 6000 K.  The data are from this paper, Boesgaard \& Tripicco
(1986), Boesgaard \& Budge (1988), and Thorburn et al. (1993) (as reanalyzed
here with the Hipparcos parameters from de Bruijne et al.~(2001)). The B
results - all detections - are the red filled squares.  The abundances are
shown on the same scale and normalized to the solar system abundances of Li
and Be (Grevesse \& Sauval 1998) as indicated by the horizontal dotted line.
There are two stars in the Li dip that appear to be deficient in B; these are
connected to their Li values by dotted lines.}
\end{figure}

\begin{figure}
\plotone{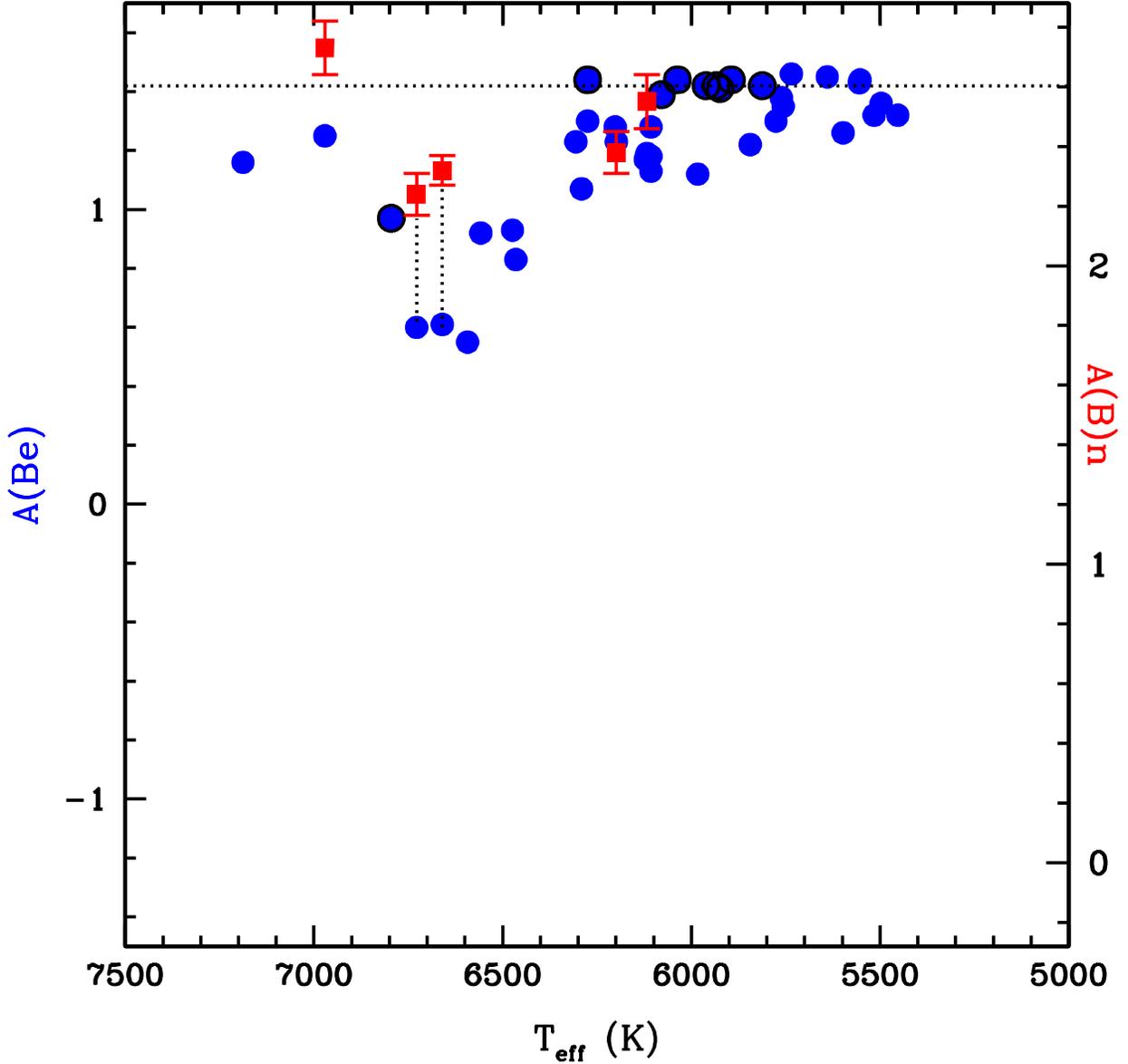}
\caption{ The Be-temperature profile for the Hyades F and G stars, on the same
vertical scale as the Li plot in Figure 10.  The blue filled circles are the
Hyades Be abundances.  The Be abundances from the Be observations reported in
this work are circled in black.  There are Be deficiencies in the same
temperature region as the Li deficiencies.  Unlike the Hyades Li abundances,
there are no Be deficiencies in the G stars.  The data are from Boesgaard \&
King (2002) (as reanalyzed in Boesgaard et al.~2004a and adjusted for the
Hipparcos parameters used here) and from the new Be observations presented
here.  The B results are shown as filled red squares on the same vertical
scale as A(Be) and as A(Li) in Figure 10.  The horizontal dotted line
corresponds to the meteoritic/solar values of Be and B.  There are two stars
in the Be dip that appear to be deficient in B; these are connected to their
Be values by dotted lines.}
\end{figure}

\begin{figure}
\plotone{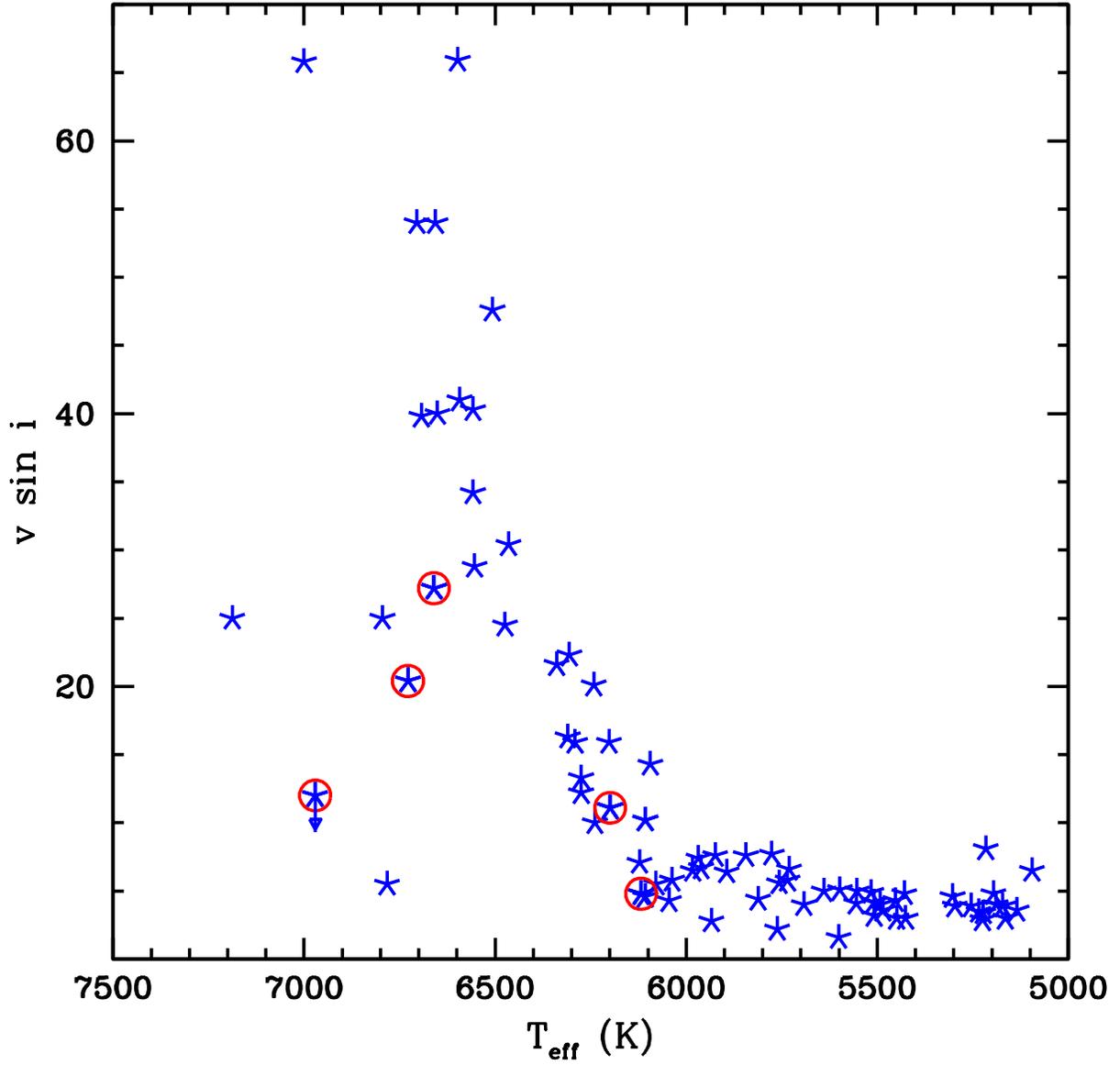}
\caption{The measured vales of v $sin$ i given in Table 3 are shown as a
function of $T_{\rm eff}$.  A dramatic drop is found from 6800 to 6200 K.  The
points for the five stars observed for B are circled in red.}
\end{figure}

\begin{figure}
\plotone{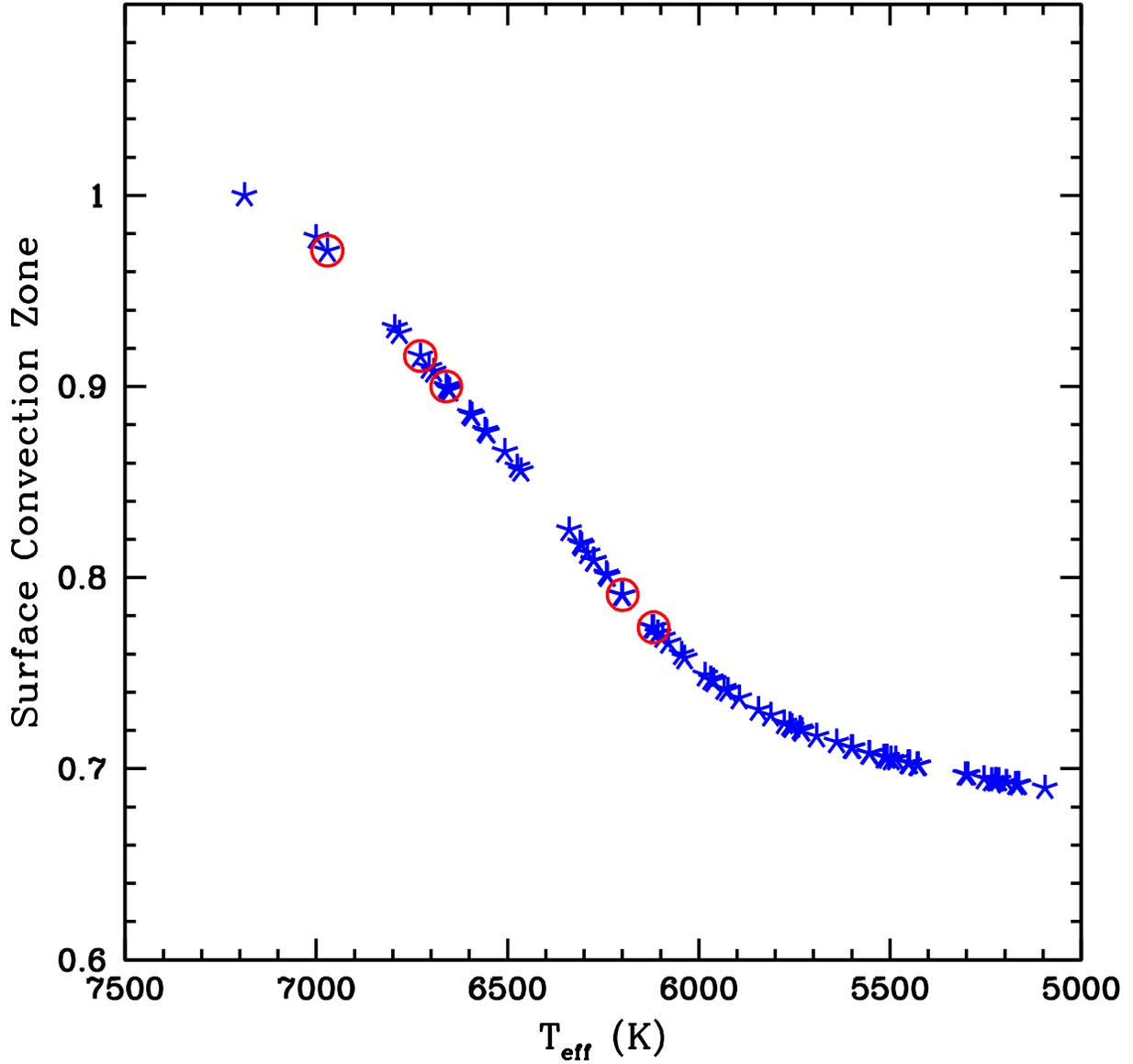}
\caption{The depth of the SCZ with respect to the fractional radius, where the
surface is 1.0 as a function of the stellar surface temperature.  The y-axis
corresponds to the boundary between the SCZ and the radiative layer.  There is
no SCZ for the hottest stars.  It deepens smoothly toward cooler stars and
reaches down to $\sim$30\% of the star by 5500 K.  The points for the five
stars observed for B are circled in red.}
\end{figure}

\begin{figure}
\plotone{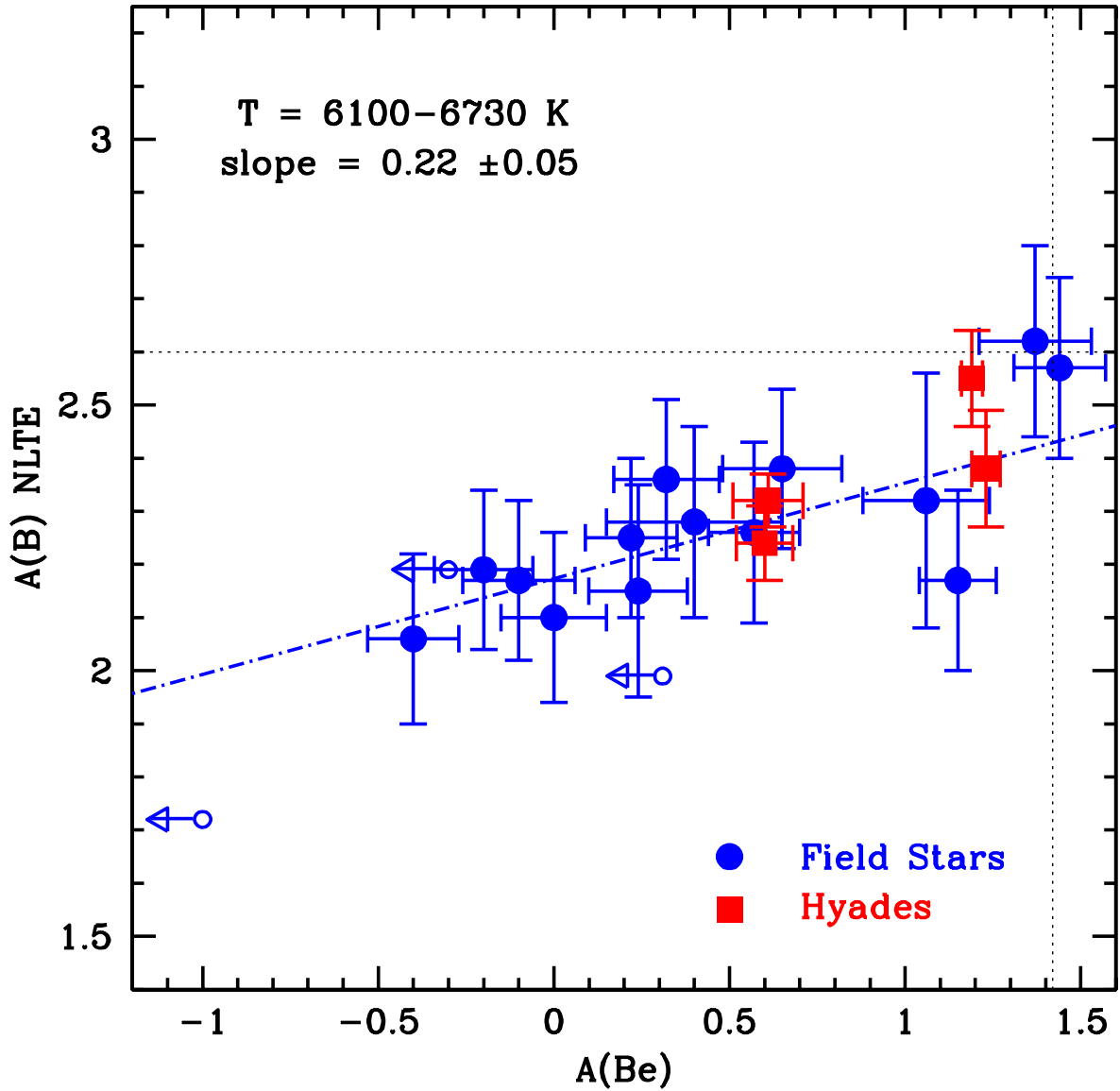}
\caption{The correlation of Be and B abundances in stars on the cool side of
the Li-Be dip. The field star data are from Boesgaard et al.~(2005).  The four
cool Hyades stars fit very well with the correlation from the field stars.
The dotted black lines show the initial Be and B abundances.}
\end{figure}

\begin{figure}
\plotone{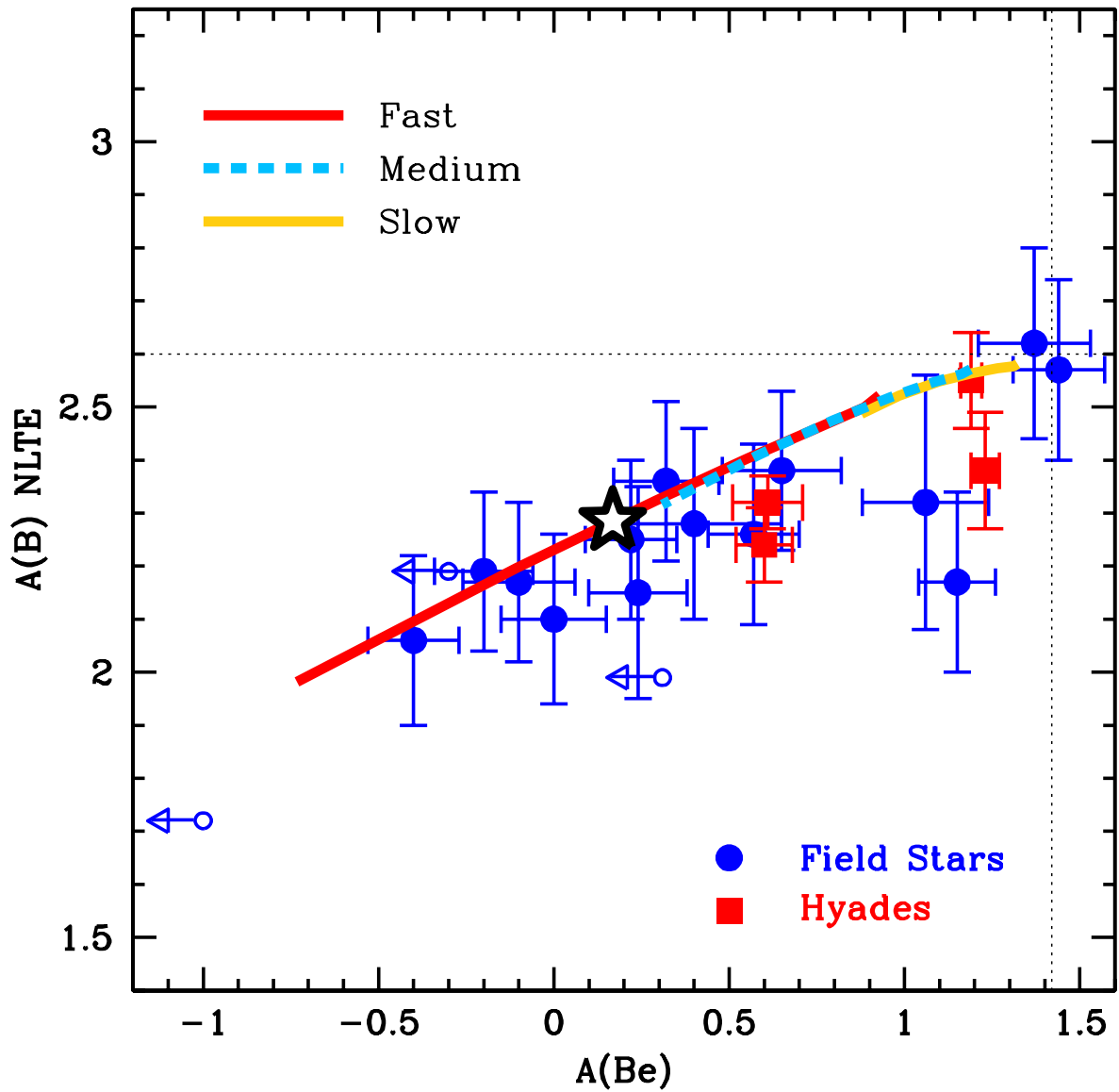}
\caption{Data points same as Figure 14, with fast, medium, and slow rotational
mixing models run to 2 Gyr over-plotted. The dotted black lines show the
initial Be and B abundances of the mixing calculations. The models reproduce
well the slope of the Be-B correlation, though are more rich in B by about 0.1
dex compared to the Hyades stars. The black star marker reflects the most
depleted model at 700 Myr, suggesting sufficient B depletion can be driven by
rotational mixing at the age of the Hyades.}
\end{figure}

\end{document}